\documentclass[preprint,12pt]{elsarticle}

\usepackage{amssymb}
\usepackage{amsmath,amsfonts}
\usepackage{algorithmic}
\usepackage{algorithm}
\usepackage{array}
\usepackage[caption=false,font=normalsize,labelfont=sf,textfont=sf]{subfig}
\usepackage{textcomp}
\usepackage{stfloats}
\usepackage{url}
\usepackage{verbatim}
\usepackage{graphicx}
\usepackage{amsmath}
\newtheorem{theorem}{Theorem}[section]

\newtheorem{definition}{Definition}[section]
\usepackage[section]{placeins}
\usepackage{color}
\usepackage{float}

\journal{ }

\begin{document}

\begin{frontmatter}

\title{Periodic Implicit Representation, Design and Optimization of Porous Structures Using Periodic B-splines}

\author[1]{Depeng Gao}
\author[1]{Yang Gao}
\author[1]{Hongwei Lin\corref{cor1}}

\affiliation[1]{organization={School of Mathematical Sciences, Zhejiang University},
            addressline={No.866, Yuhangtang Rd}, 
            city={Hangzhou},
            postcode={310058}, 
            state={Zhejiang Provence},
            country={China}}

\cortext[cor1]{Corresponding author. E-mail address: hwlin@zju.edu.cn (H. Lin).}

\begin{abstract}
Porous structures are intricate solid materials with numerous small pores, extensively used in fields like medicine, chemical engineering, and aerospace. 
    However, the design of such structures using computer-aided tools is a time-consuming and tedious process.
    In this study, we propose a novel representation method and design approach for porous units that can be infinitely spliced to form a porous structure. 
    We use periodic B-spline functions to represent periodic or symmetric porous units. 
    Starting from a voxel representation of a porous sample, the discrete distance field is computed. 
    To fit the discrete distance field with a periodic B-spline, we introduce the constrained least squares progressive-iterative approximation algorithm, which results in an implicit porous unit. 
    This unit can be subject to optimization to enhance connectivity and utilized for topology optimization, thereby improving the model's stiffness while maintaining periodicity or symmetry. 
    The experimental results demonstrate the potential of the designed complex porous units in enhancing the mechanical performance of the model. 
    Consequently, this study has the potential to incorporate remarkable structures derived from artificial design or nature into the design of high-performing models, showing the promise for biomimetic applications.
\end{abstract}

\begin{graphicalabstract}
\includegraphics[width=1.0\textwidth]{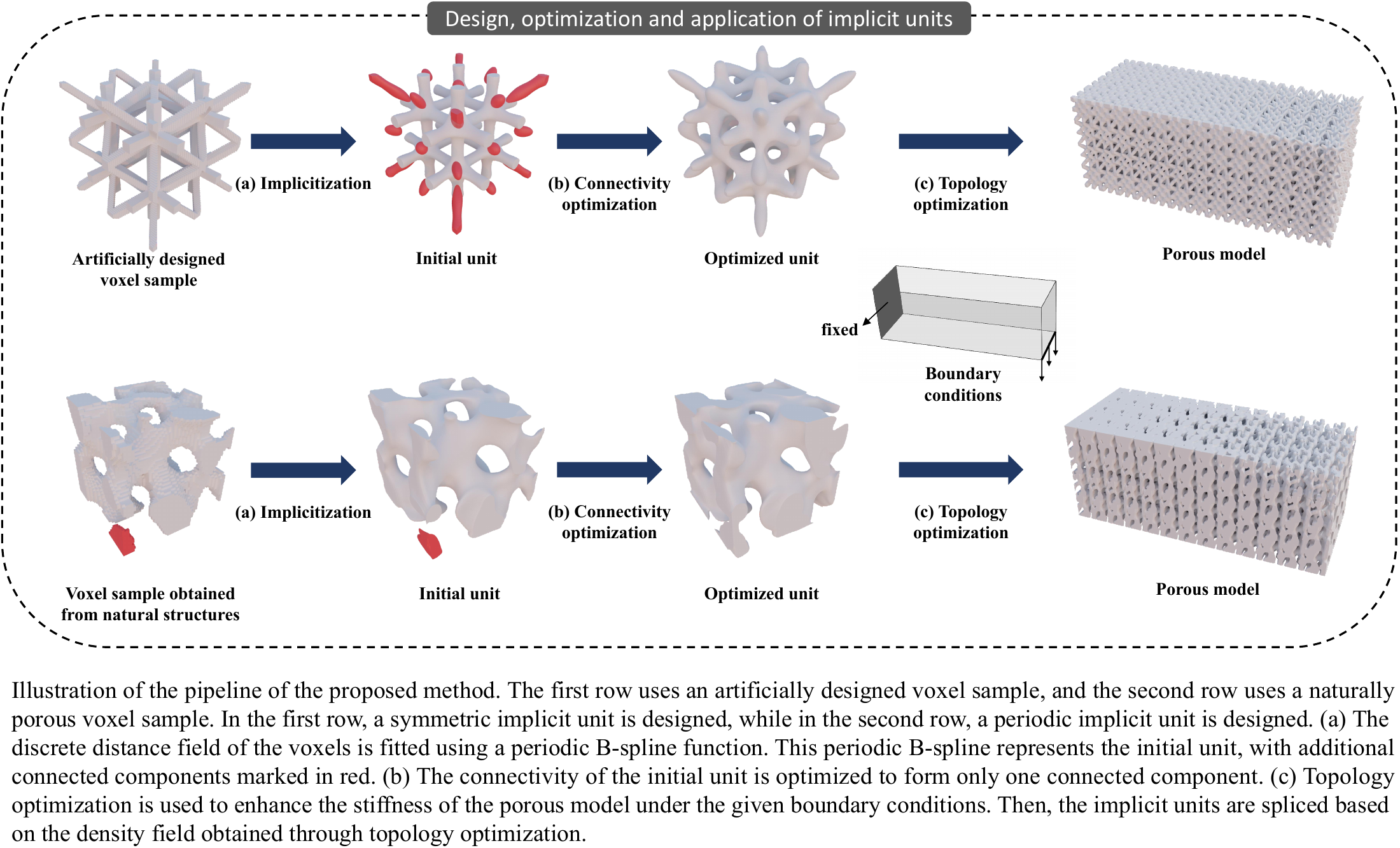}
\end{graphicalabstract}

\begin{highlights}
    \item We introduce the periodic B-spline function as a novel representation for periodic or symmetric porous units.
    \item We propose the CON-LSPIA algorithm, which solves constrained fitting problems encountered during the design of implicit porous units. 
    \item The new representation maintains the periodicity or symmetry of porous units during connectivity optimization.
    \item Implicit units under the new representation can be combined with homogenization methods to reduce the computational cost of topology optimization.
\end{highlights}


\end{frontmatter}

\section{Introduction}
Porous structures are widely used in the aerospace, bone tissue engineering, and chemical engineering fields due to their high specific strength, lightweight, and large specific surface area. 
    The advent of additive manufacturing (AM) has enabled the fabrication of complex porous structures. 
    Rather than designing complete porous structures, researchers prefer to design porous units that can be spliced to form a porous structure. 
    Designing a porous unit is easier due to its smaller size compared to a complete porous structure. 
    In addition, considering the porous unit as a novel material can significantly reduce the computational costs of simulation analysis. 
    Several studies~\cite{al2018microarchitected, zhang2021mechanical} have shown that the morphology of the units significantly affects the mechanical properties of the spliced porous structure.
    Therefore, the design of porous units is an important issue.
    However, commercial CAD software has limitations in designing porous units with complex topological and geometric structures~\cite{savio2018geometric}.
    Therefore, it is a worthwhile research issue to explore a simplified representation method and propose a design approach based on this representation for porous units.
    
In recent years, there has been an increasing interest in the functional representation of porous units.  
    A porous unit can be represented by a simple formula, called an implicit unit. 
    The use of R-functions~\cite{gao2023free} allows Boolean operations on implicit units while maintaining continuity~\cite{pasko2011procedural}. 
    Meanwhile, implicit units can be directly manufactured through AM without the need for conversion to triangular meshes~\cite{hong2023implicit,yan2023adaptive}.
    Among the various implicit units, one of the most popular ones is the triply periodic minimal surface (TPMS)~\cite{feng2022triply}. 
    TPMSs are periodic and can be infinitely spliced in three directions, making them suitable for a wide range of applications~\cite{feng2022triply}. 
    However, TPMSs have some limitations. 
    The algebraic equations have a limited number of adjustable parameters, which limits the optimization space for meeting different application requirements, such as density, wall thickness, and connectivity. 
    Additionally, the limited types of existing TPMSs hinder performance in multi-unit topology optimization~\cite{feng2022stiffness}. 
    
 Another type of implicit unit is artificially designed units. These units are often derived artificially due to the challenges of interactive design for porous structures under implicit representation~\cite{pasko2011procedural}. Typically, artificially designed units have simpler structures compared to natural porous units. 
    Different from direct derivation, Gao et al.~\cite{gao2022connectivity} use B-spline functions to represent porous structures and perform inverse design on these structures. 
    Although B-spline functions can represent complex porous structures as intricate as natural ones, it becomes challenging to use them for large-sized structures. 
    As the size of the porous structure increases, more control coefficients and computation time are required for an accurate representation.
    Simulating large-scale porous structures using the finite element method also adds to the computational burden due to the necessity of numerous meshes.
    However, these challenges can be addressed by employing periodic functions to represent complex porous structures. 
    By combining periodic functions with translation operations, we can represent porous structures of arbitrary sizes. 
    Additionally, this representation method is particularly advantageous when dealing with porous models that consist of repetitive structures since it significantly reduces the computational effort of simulation through homogenization methods~\cite{dong2019149}. 
    Thus, the introduction of periodicity is necessary to facilitate the representation and simulation of large-scale porous models.

To the best of our knowledge, the currently available artificial implicit units can not simultaneously meet all the necessary requirements. These requirements include controllable density, the ability to form connected structures upon splicing, usability for simulation analysis and topology optimization, and ease of design. 
    To tackle these problems, we take a similar approach to the one used in~\cite{gao2022connectivity}, that is, obtaining an implicit porous structure by fitting a discrete distance field. 
    The functions obtained from fitting can represent intricate porous units, resembling structures found in nature. 
    In contrast to~\cite{gao2022connectivity} where a fixed-size porous structure is designed, we generate a porous unit and then splice it to form a porous structure of arbitrary size. 
    These porous units can be regarded as novel materials for simulation analysis. 
    Moreover, the density of the units can be adjusted, making them suitable for applications such as topology optimization. 
    However, the main challenge lies in acquiring and representing units that can be infinitely spliced. 
    Additionally, the problem of disconnected units frequently arises in reverse design. 
    Maintaining the periodicity of the units during the connectivity optimization process is also challenging. 
    
In this study, we address these issues by utilizing periodic B-splines as a new representation for implicit porous units. 
    This representation can effectively represent both periodic units, which can be infinitely spliced in space, and symmetric units, which not only can be infinitely spliced but also have symmetric structures within the unit itself. 
    Furthermore, we propose a design method corresponding to this representation. 
    Specifically, we start with a porous sample represented by voxels obtained from Computed Tomography (CT) or artificially designed. Then a periodic B-spline is generated based on this sample. 
    Subsequently, the discrete distance field of the voxels is calculated using the Distance to Measure (DTM) function. 
    Afterward, the discrete distance field is fitted by a periodic B-spline function using the constrained least squares progressive-iterative approximation (CON-LSPIA) algorithm. 
    Finally, the implicit porous unit represented by the fitted periodic B-spline can be optimized to form only one connected component after splicing and then utilized for topology optimization. 
    In summary, the main contributions of this study are as follows: 
    \begin{itemize}
        \item We introduce the periodic B-spline function as a novel representation for periodic or symmetric porous units.
        \item We propose the CON-LSPIA algorithm, which solves constrained fitting problems encountered during the design of implicit porous units. 
        \item The new representation maintains the periodicity or symmetry of porous units during connectivity optimization.
        \item Implicit units under the new representation can be combined with homogenization methods to reduce the computational cost of topology optimization.
    \end{itemize}

The remainder of this paper is organized as follows. 
    First, in Subsection~\ref{subsec: Related work}, we provide a review of the existing literature on the representation of porous structures. 
    In Section~\ref{sec: Preliminaries}, we introduce the preliminaries on trivariate B-spline functions and implicit representations of porous structures.
    Next, in Section~\ref{sec: Method}, we present the definition and generation process of periodic B-splines.     
    Following that, in Section~\ref{sec: Discussion}, we discuss the design, connectivity optimization, and topology optimization of the porous units represented by periodic B-splines. 
    Furthermore, we illustrate the effectiveness of the developed method through several experimental examples in Section~\ref{sec: Implementation, results and discussions}.
    Finally, we conclude this study in Section~\ref{sec: Conclusion}.

\subsection{Related work}
\label{subsec: Related work}
The design of porous units with complex geometric and topological structures is challenging for CAD modeling systems. 
    Recent research has focused on proposing simpler representation and design methods for porous units. 
    From a geometric perspective, the representation and design of porous units can be categorized into four types: boundary representation (B-rep), parameter representation, volume representation (V-rep), and functional representation (F-rep). 
    An overview of these methods is given in this section.

\textbf{B-rep} is a widely used method in geometric modeling. 
    Porous units can be represented by polygonal meshes from a discrete perspective. 
    Directly design of a porous unit consisting of a large number of facets is challenging. 
    Parametric surfaces can be used to represent porous units from a continuous perspective. 
    Massarwi et al.~\cite{massarwi2018hierarchical} use B-spline surfaces to represent a porous unit.
    Due to the topological complexity of porous units, the representation of a porous unit typically requires the joining of multiple B-spline surfaces. 
    Vongbunyong et al.~\cite{vongbunyong2017rapid} propose a hybrid B-rep and polygon approach to reduce the computational cost of intersections when splicing B-rep units. The resulting spliced porous structure can be directly converted into an STL file.
    Intersections are inevitable during the splicing and design of B-rep units, and solving intersection problems caused by the design and splicing of complex structures remains a challenge in B-rep.

\textbf{Parameter representation} refers to the use of a set of parameters to describe the geometric and topological structure of porous units. 
    Body-center Cubic (BCC)~\cite{ren2019multi} and Face-centered Cubic (FCC)~\cite{li2019novel} units are widely used parameter representations of porous units. 
    These units are controlled by various parameters, including rod distribution, rod shape, and cross-sectional area. 
    By optimizing the parameters, the optimal unit can be obtained from a parameterized unit library based on the desired objective, such as stiffness~\cite{liu2020rapid,li2020anisotropic}. 
    The units can be input into neural networks as they can be described by a set of parameters. 
    Wang et al.~\cite{wang2022machine} optimize the unit parameters using machine learning methods to maximize the stiffness of porous models. 
    In the literature~\cite{abu2023inverse}, researchers trained a network to obtain the desired parameterized unit that meets the expected physical performance. 
    The complexity and size of the parameterized unit library generally depend on the number of meaningful parameters. 
    Typically, parameterized units have few parameters and simple structures due to the fact that the artificially designed parameters should have geometric meaning, such as angle, cross-sectional area, and position.
    Although parameterized units have shown excellent mechanical performance in practical applications, their simple structures may not be suitable for clinical requirements~\cite{chen2020porous}. 
    Therefore, a larger library of parameterized units is needed to expand the solution space for application problems.

Porous units under \textbf{V-rep} can be categorized into discrete and continuous representations.
    Trivariate B-spline solids are commonly used for the continuous representation of porous units~\cite{massarwi2018hierarchical,hong2021conformal}. 
    The advantage of representing porous units using B-spline solids is that they can be used directly for simulation analysis. 
    However, further research is needed to design complex porous units using trivariate B-spline solids and to control the attributes (porosity, pore size, and wall thickness) of porous units. 
    
Porous units are commonly represented using voxels in a discrete representation.
    A voxelized porous unit can be considered as a binary image, where the solid voxels are labeled 1 and the void voxels are labeled 0.
    As a result, compared to B-Rep, the time consumption and errors associated with voxel intersections are greatly reduced by using logical operations. 
    Taking advantage of this, Aremu et al.~\cite{aremu2017voxel} proposed a design method for complex porous models based on voxel representation. 
    Another advantage of voxel representation is its suitability for reverse design. 
    The development of CT technology has enabled the acquisition of higher-resolution images of porous structures \cite{feng2020end}. 
    Several statistical methods have been proposed to reconstruct porous structures of equal or larger size from these high-resolution images~\cite{holdstein2009volumetric,hasanabadi2022two,li2019dictionary}.
    In recent years, methods based on deep learning have been proposed to reverse design porous structures from porous samples~\cite{feng2020end,zhang2021slice}. 
    Although both forward and reverse design methods are suitable for designing complex porous structures using voxel representation, there are several limitations that restrict their application. 
    First, the size of the porous structure is fixed, requiring a larger number of voxels to represent larger structures. 
    Second, reverse-designed structures are non-periodic, requiring a significant amount of time to synthesize large porous structures compared to generating periodic units and splicing them together. 
    Finally, the voxel representation is not smooth, leading to problems such as stress concentration.

In \textbf{F-rep} models, a porous unit can be represented by a function called an implicit unit. 
    TPMSs are the most widely applied F-rep units. A TPMS solid can be represented as $\phi_{TPMS} \leq c$, where $\phi_{TPMS}$ is a periodic function in $\mathbb{R}^3$, and $c$ is called threshold~\cite{hu2021heterogeneous}. 
    The threshold can control the relative density, isotropy, wall thickness, and other properties of TPMS units \cite{feng2021isotropic}. 
    Moreover, model strength can be improved by simultaneously optimizing the period and wall thickness of TPMSs~\cite{yan2019strong}. 
    Multi-unit topology optimization is a recent research hotspot that aims to assign different types of units with unique properties to appropriate positions in the design space to achieve optimal mechanical performance~\cite{feng2022stiffness,ozdemir2023novel,shi2021design}. 
    However, the solution space for optimizing porous models is limited due to the few types of TPMS units, and the control of TPMSs can only be achieved through thresholds and periods. 

Since F-reps can represent topologically complex porous units, the artificial design of implicit units is a potential method to expand the current database of complex porous units. 
    Pasko et al.~\cite{pasko2011procedural} generated a porous structure by splicing a given implicit unit through the composition of periodic functions. Subsequently, the R-function is utilized to generate porous models. 
    Hong et al.~\cite{hong2023implicit} constructed implicit porous structures by designing curves and fitting the corresponding distance fields. 
    Because the implicit units in their work are artificially designed, they are relatively simple compared to natural porous structures. 
    Massarwi et al.~\cite{massarwi2018hierarchical} proposed a method to design randomly generated implicit porous structures by imposing constraints on the unit boundaries to ensure connectivity. 
    However, the relative density of the units is difficult to control due to their randomness. 
    Gao et al.~\cite{gao2022connectivity} represent porous structures using implicit B-spline functions. The implicit porous structures are obtained by fitting the distance field of voxels synthesized by 3D texture synthesis from porous samples. 
    Although the proposed method can generate complex implicit structures, it shares the same drawback of fixed size as voxel-based representation and is also difficult to perform simulation analysis. 
    Although F-rep can represent complex structures, the artificially designed units are simpler than natural porous structures and do not fully utilize the advantages of F-rep. 
    Additionally, the effectiveness of artificially designed implicit units has not been evaluated through simulation analysis and other tests. 
    Moreover, implicit models are often obtained by implicitizing other representations, errors are introduced in this process. 
    As mentioned in~\cite{feng2018review}, there is a need for an effective method to design complex porous units suitable for analysis.  

In summary, voxel representation has the advantage of easy interactive design, parameter representation is easy to control and optimize, and F-rep can represent complex geometric models. 
    Therefore, based on these considerations, the objective of this study is to introduce a novel representation and an associated design method that combines the aforementioned benefits. 

    \section{Preliminaries}
    \label{sec: Preliminaries}
    \subsection{Trivariate B-spline function}
    \label{subsec:Trivariate B-spline solid}
    A B-spline function $C(u)$ with degree of $p$ is defined by:
    \begin{equation}
        C(u) = \sum_{i=0}^{n-1}N_{i,p}{C_i} \quad 0\leq u \leq 1,
    \end{equation} 
    where the $C_i$ is the $i$-th control coefficient, and the $N_{i,p}(u)$ is the B-spline basis with degree of $p$, which is defined on the knot vector~\cite{piegl1996nurbs}:
    \begin{equation} 
        \mathbf{T}=\{\ \underbrace{t_0,\cdots,t_p}_{0},\cdots, \underbrace{t_{m-p-1},\cdots,t_{m-1}}_{1} \}.
    \end{equation}
    
    A trivariate B-spline function with a degree of $(p_u,p_v,p_w)$ in each direction is a tensor product volume defined within its parametric domain $[0,1]\times [0,1] \times [0,1]$:
    \begin{equation}
        \begin{aligned}
             & C(u,v,w) = \sum_{i=0}^{n_u-1}\sum_{j=0}^{n_v-1}\sum_{k=0}^{n_w-1}R_{ijk}(u,v,w)C_{ijk} \\
        & R_{ijk}(u,v,w) = N_{i,p_u}(u)N_{j,p_v}(v)N_{k,p_w}(w), 
        \end{aligned}
    \end{equation}
    where $C_{ijk} \in \mathbb{R}$ is the control coefficient, $N_{i,p_u}(u)$, $N_{j,p_v}(v)$ and $N_{k,p_w}(w)$ are the B-spline basis in each direction, $R_{i,j,k}(u,v,w)$ is the blending basis function, and, $n_u$, $n_v$ and $n_w$ are the number of control coefficients in three direction. 
    
    \subsection{Implicit representation of porous structure}
    \label{subsec:Construction of discrete distance field of porous model}
    Typically, a solid implicit porous unit can be represented in the following three forms:
    \begin{equation}
        \begin{aligned}
            \label{eq: porous_representation}
            &\phi(u,v,w) \leq c(u,v,w)  \\
            &\phi(u,v,w) \geq c(u,v,w)   \\ 
            &c_1(u,v,w) \leq \phi(u,v,w) \leq c_2(u,v,w) \quad (u,v,w) \in \mathcal{D},
        \end{aligned}
    \end{equation}
    where $\phi$ is a function represents a porous unit, $c$, $c_1$, and $c_2$ are threshold distribution fields that control the range of extraction from the function $\phi$, and $\mathcal{D}$ refers to the region where the unit exists. 
        In this study, we assume that a solid implicit porous unit is represented as $\phi(u,v,w) \leq c(u,v,w)$, because the second and third forms can be equivalent to the first form, and $\mathcal{D}= [0,1]^3$.  
    
    The porous units can be spliced to generate a porous structure with an arbitrary size. 
    Assuming that the porous structure consists of $E_u$, $E_v$, and $E_w$ units in each direction. 
        To simplify the design process, we design a porous structure within a unit cube and then scale it in each direction to achieve a desired volume of $E_u\times E_v\times E_w$. 
        The implicit porous units are reduced and spliced within the unit cube to construct a porous structure. 
        The units are spliced using the following transformation:
    \begin{equation}
        \label{eq:transformation}
        \psi(E_u,u) = uE_u-\lfloor uE_u\rfloor. 
    \end{equation}
    
    For simplicity, we denote $\psi(E_u,u)$ as $\psi(u)$, and apply the same treatment to $\psi(v)$ and $\psi(w)$. The porous structure $\varphi$ within the unit cube is defined as follows:
    \begin{equation}
        \label{eq:implicit porous structure}
        \varphi(u,v,w) = \phi(\psi(u),\psi(v),\psi(w)).
    \end{equation}  
    
    To control the density of the porous structure and create a solid structure, the porous structure is defined as $\varphi(u,v,w) \leq c(u,v,w)$, where $c(u,v,w)$ is the threshold distribution field. 
        In this study, the threshold distribution field is represented by a trivariate B-spline function. 

    \section{Definition and generation of periodic B-spline}
    \label{sec: Method}
    \begin{figure*}[h]
    \centering
    \includegraphics[width=0.95\textwidth]{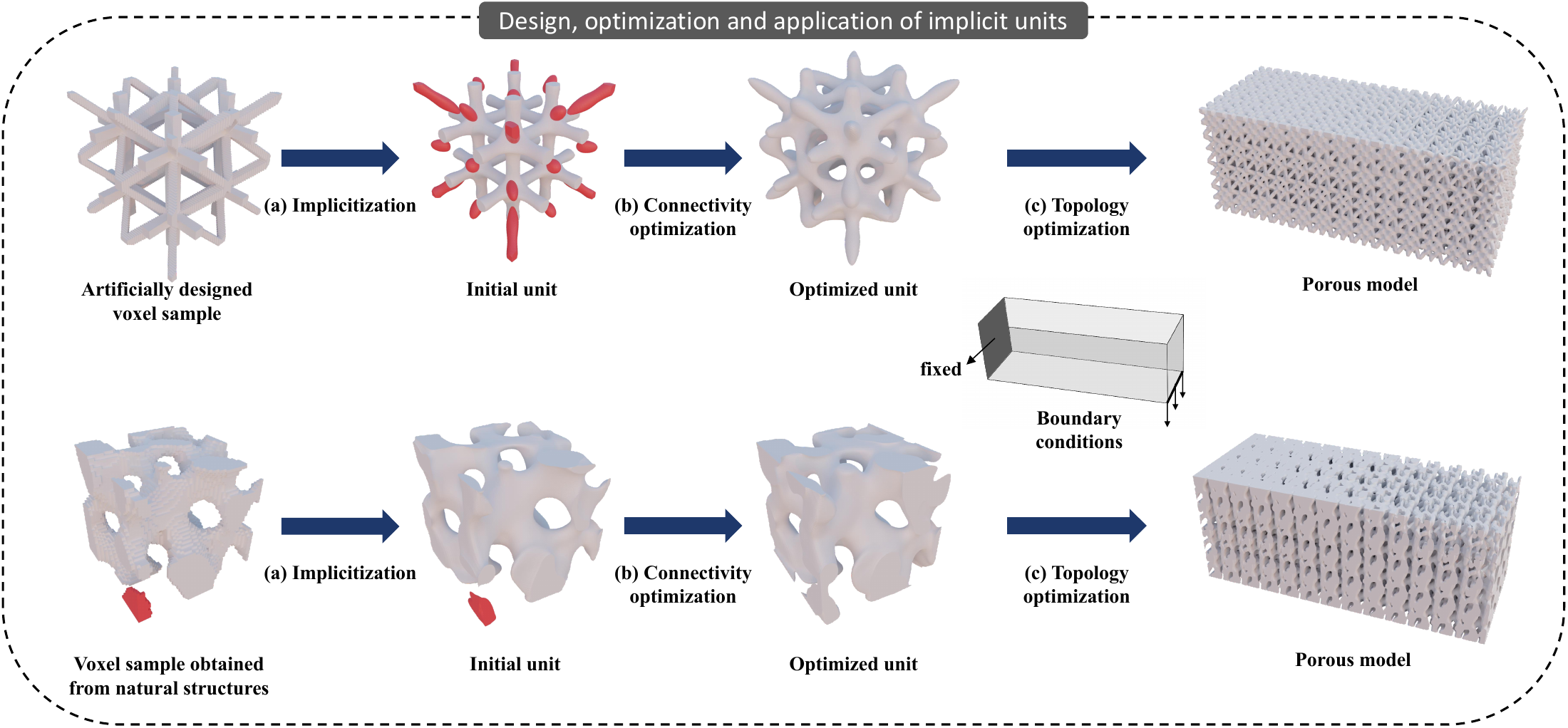}
    \caption{Illustration of the pipeline of the proposed method. The first row uses an artificially designed voxel sample, and the second row uses a naturally porous voxel sample. In the first row, a symmetric implicit unit is designed, while in the second row, a periodic implicit unit is designed. (a) The discrete distance field of the voxels is fitted using a periodic B-spline function. This periodic B-spline represents the initial unit, with additional connected components marked in red. (b) The connectivity of the initial unit is optimized to form only one connected component. (c) Topology optimization is used to enhance the stiffness of the porous model under the given boundary conditions. Then, the implicit units are spliced based on the density field obtained through topology optimization.} 
    \label{fig:flow_chart}
    \end{figure*}
    
    The pipeline for the proposed design methods of implicit porous units is illustrated in Figure~\ref{fig:flow_chart}. 
        Considering a porous sample represented by voxels, this method transforms the porous sample into an implicit representation, where the implicit porous units exhibit periodic or symmetric behavior. 
        The discrete distance field of the porous sample is first calculated using the DTM function. 
        Then, the discrete distance field is fitted by a periodic B-spline function using the proposed CON-LSPIA method to generate an initial implicit unit. 
        Additionally, the implicit unit can be optimized to form only one connected component while keeping symmetry or periodicity. 
        Subsequently, given a boundary condition, the optimal density distribution of the implicit unit is determined using topology optimization. 
        Finally, the porous units can be spliced to obtain a porous model that achieves the desired density field. 
    \subsection{Definition and property of periodic B-spline}
    \label{subsec: Periodic B-spline}
    Ensuring the periodicity of the implicit porous units is essential for constructing implicit porous structures. 
        Additionally, certain excellent units, like P-TPMS and IWP-TPMS, exhibit symmetry~\cite{feng2022triply}. 
        As demonstrated in~\cite{pasko2011procedural}, various periodic functions can be used to represent a periodic porous unit. 
        However, there are limited functions available to represent symmetric porous units. 
        Furthermore, the optimization and adjustment of implicit units are necessary to meet the practical application requirements. 
        However, preserving the periodicity or symmetry of the porous units during the optimization process presents a challenge. 
        To accurately represent structures with periodicity or symmetry, the periodic B-spline is employed as a novel representation.
        
    By assigning periodicity to knot vectors and control coefficients, a periodic B-spline can be defined as follows:
    \begin{definition}
        \label{periodic B-spline}
     A \textit{periodic B-spline} is a B-spline where the first $k$ control points are the same as the last $k$ control points, and the first $k$ knot intervals in the knot vector are equivalent to the last $k$ intervals~\cite{barnhill2014computer}.
    \end{definition}
    
    \begin{definition}
        \label{def:symmetric degree}
        Let $C(u)=\sum_{i=0}^{n-1}N_{i,p}(u)C_i$ be a B-spline function with degree $p$ and corresponding knot vector:
        \begin{equation*}
            \mathbf{T}=\{\ \underbrace{t_0,\cdots,t_p}_{0},\cdots, \underbrace{t_{m-p-1},\cdots,t_{m-1}}_{1} \}.
        \end{equation*}
        We call this B-spline a \textit{periodic B-spline function with a symmetric degree of r} when it meets the following conditions:
    \begin{enumerate}
        \item $t_{i+1}-t_{i} = t_{m-i-1}-t_{m-i-2}, \quad 0 \leq i < r+p.$
        \item $C_i = C_{n-1-i}, \quad 0\leq i < r $.
        \item {$0 \leq r \leq n$}.
    \end{enumerate}
    \end{definition}
    
    It can be inferred from this definition that as the symmetric degree $r$ increases, the size of the symmetric region in the B-spline function also becomes larger. 
        As proven in the Supplementary Material, this discovery can be attributed to the following theorem:
    
    \begin{theorem}
    \label{theorem: periodic B-spline}
    A B-spline function $C(u)$ with a symmetric degree of $r$ and a knot vector $\mathbf{T}$ satisfies the following equation: 
    \begin{equation}
    \label{eq:B-spline curve symmetry}
        C(u)=C(1-u),\quad u \in [0,t_r)
    \end{equation}
    Moreover, when $r = \lfloor {\frac{n}{2}}\rfloor$, the implicit B-spline is symmetric in $[0,1]$, that meets the following formula:
    \begin{equation}
        C(u)=C(1-u),\quad u \in [0,1].
    \end{equation}
    \end{theorem}
    The Theorem~\ref{theorem: periodic B-spline} can be naturally extended to a trivariate B-spline function.
    
    \begin{theorem}
    A trivariate B-spline function $C(u,v,w)$ of degree $\mathbf{p}=(p_u,p_v,p_w)$ with a symmetric degree of $\mathbf{r}=(r_u,r_v,r_w)$ satisfies the equation:
    \begin{equation}
    \begin{aligned}
         C(u,v,w)&=C(1-u,v,w)=C(u,1-v,w)\\
         &=C(u,v,1-w)
    \end{aligned}
    \end{equation}
    when $(u,v,w)\in [0,t_{r_u}^u]\times[0,t_{r_v}^v]\times[0,t_{r_w}^w]$, where $t_{r_u}^u$ is the $r_u$-th knot in the u-direction, and $t_{r_v}^v$ and $t_{r_w}^w$ are defined in the same way. Moreover, the implicit B-spline is symmetric in $[0,1]\times[0,1]\times[0,1]$ when $r_u = \lfloor {\frac{n_u}{2}} \rfloor$, $r_v = \lfloor {\frac{n_v}{2}} \rfloor$, and $r_w = \lfloor {\frac{n_w}{2}} \rfloor$, where $n_u\times n_v \times n_w$ is the number of control coefficients.
    \end{theorem}
    
    In conclusion, a trivariate B-spline function with symmetric degree of $\mathbf{r}=(r_u,r_v,r_w)$ can represent a periodic porous unit, given that $r_u,r_v,r_w > 0$. 
        Moreover, the trivariate B-spline function can represent a symmetric porous unit if $r_u = \lfloor {\frac{n_u}{2}} \rfloor$, $r_v = \lfloor {\frac{n_v}{2}} \rfloor$, and $r_w = \lfloor {\frac{n_w}{2}} \rfloor$.
    
    \subsection{Construction of discrete distance field}
    \label{subsec:Construction of discrete distance field}
    Section~\ref{subsec: Periodic B-spline} introduces the utilization of the periodic B-spline function to represent the periodic or symmetric porous unit. 
        In this subsection, the periodic B-spline is constructed by fitting a discrete distance field of a porous sample. 
        Assuming that the porous samples are represented by voxels (which are equal to binary images) obtained from artificially designed or three-dimensional imaging techniques such as CT. 
        
\begin{figure} [h]
        \centering
        \subfloat[]{
            \includegraphics[width=0.20\textwidth]{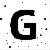}}
        \subfloat[]{
            \includegraphics[width=0.25\textwidth]{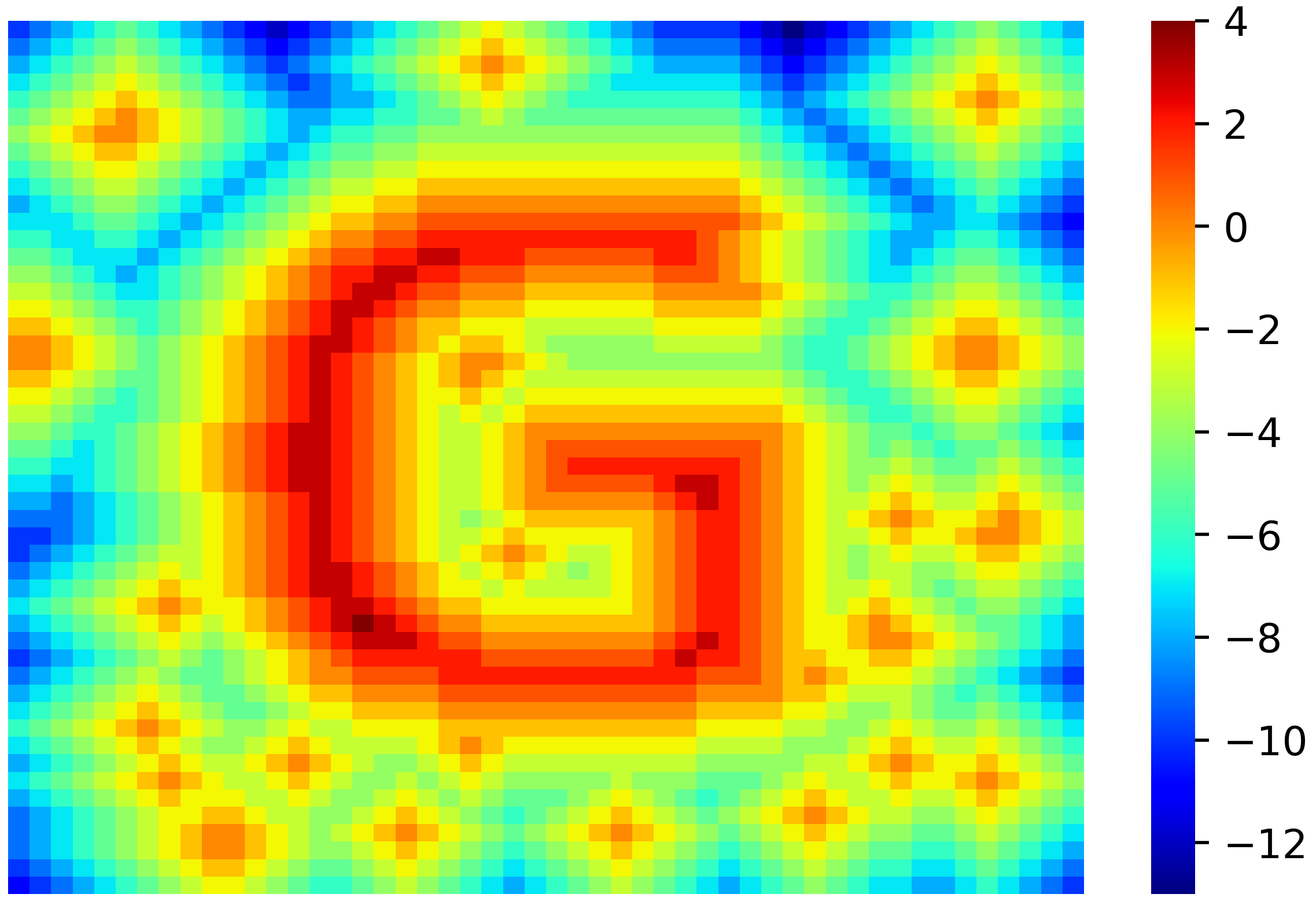}}
        \subfloat[]{
            \includegraphics[width=0.25\textwidth]{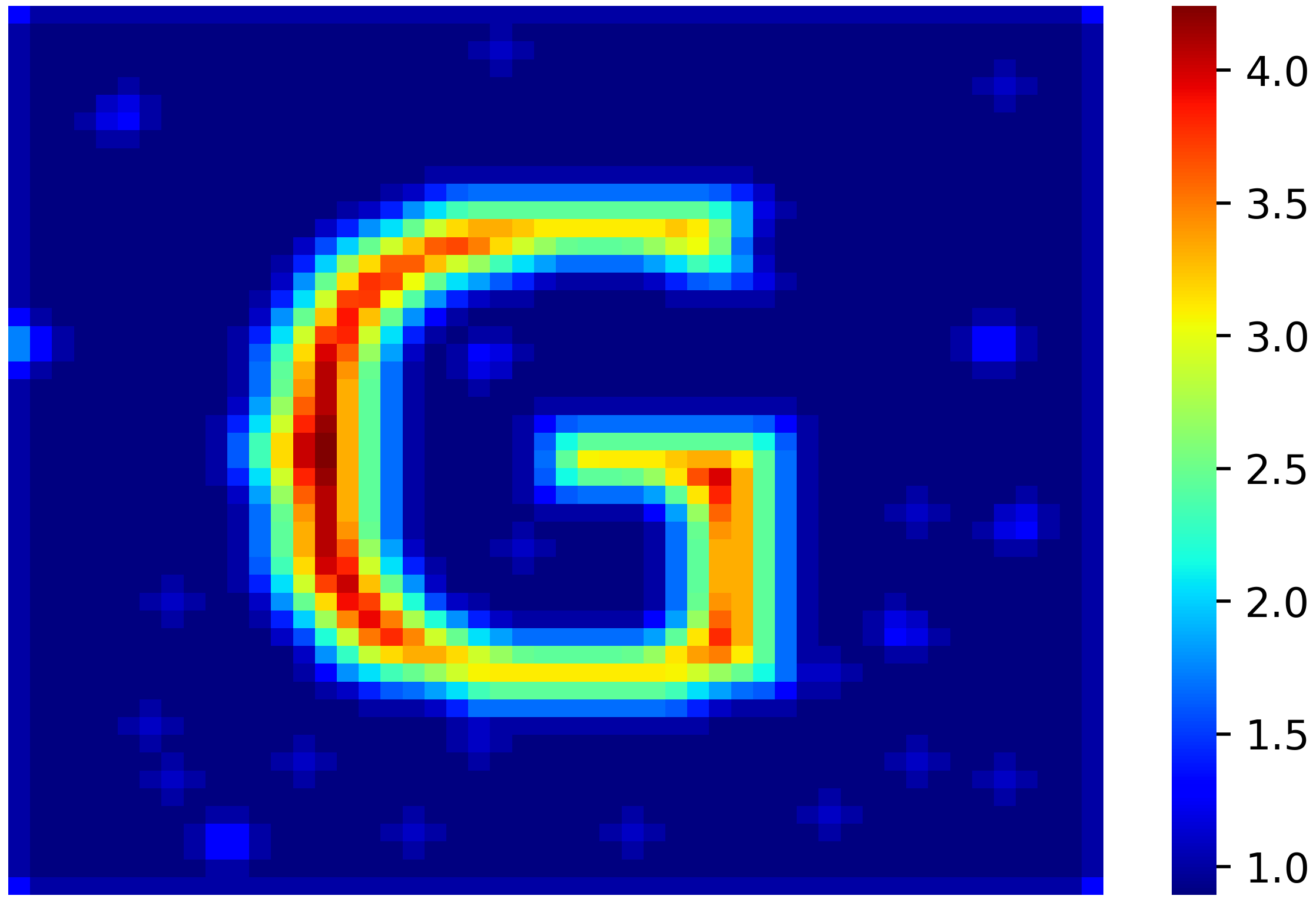}}
        \caption{Comparison of several methods for calculating discrete distance fields. (a) A 2D binary image. (b) Manhattan distance field. (c) DTM field with $m=5$}
        \label{fig:distance_field} 
     \vspace{-0.3cm}
    \end{figure}

    A high-quality discrete distance field is required to generate an accurate implicit representation of a porous unit. 
       However, there may be noise in the 3D image that affects the quality of the distance field. 
       As illustrated in Figure~\ref{fig:distance_field}(a), we artificially introduce noise while constructing a letter "G". 
       Figure~\ref{fig:distance_field}(b) presents a discrete distance field obtained using the Manhattan distance field~\cite{obayashi2018persistence}. 
       The implicit values for small connected components, which correspond to the noise, are observed to be as significant as those for the solid material "G". 
       This phenomenon results in a bad implicitization. 
       To overcome this challenge, the DTM function for images~\cite{gao2022connectivity,anai2020dtm} is employed to obtain a high-quality distance field that is insensitive to noise data. 
       The distance value assigned to a pixel $\mathbf{x}$ in the image is defined as follows:
    \begin{equation}
    \label{eq:dtm}
        d_m^2(\mathbf{x}) = \frac{1}{m}(\sum_{i=1}^{k-1}\omega_i\|\mathbf{x}-\mathbf{p_i}\|^2+(m-\sum_{i=1}^{k-1}\omega_i)\|\mathbf{x}-\mathbf{p_k}\|^2),
    \end{equation}
    where $\mathbf\{p_i\}_{i=1}^k$ represents the set of k-nearest pixels of $\mathbf{x}$ and $\{\omega_i\}_{i=1}^k$ denotes the corresponding pixel values (either 0 or 1) satisfying the condition $\sum_{i=1}^{k-1}\omega_i<m\leq \sum_{i=1}^k\omega_i$.
     
    As shown in Figure~\ref{fig:distance_field}(c), the image becomes blurry and the noise becomes insignificant. 
        We set $m=5$ for all subsequent experiments. 
    
    Our method is applicable to both porous samples represented by F-rep and B-rep. 
        In the case of F-reps, the discrete distance field can be sampled directly from the implicit function. 
        Moreover, previous studies have proposed various methods for converting B-reps into F-reps~\cite{baerentzen2005robust}. 
        
    Whether the discrete distance field is obtained from Frep, Brep, or voxels, it can be considered as a grid where each cell center corresponds to a value. 
        Consequently, we can obtain a series of grid center positions $\{(u_s,v_s,w_s)\}_{s=0}^{S-1}$ and their corresponding values $\{X_s\}_{s=0}^{S-1}$ as data points. 
        These data points are fitted using a B-spline function to create an implicit unit. 
        Constraints are necessary to ensure the periodicity or symmetry of the implicit representation. 
        In the following section, an unconstrained fitting problem will be modeled to approximate these data points using periodic B-splines. 
    
    \subsection{Constrained-LSPIA}
    \label{subsec: fitting}
    Section~\ref{subsec: Periodic B-spline} illustrates that a periodic B-spline can represent a periodic or symmetric porous unit. 
        Additionally, the periodic B-spline can be obtained by fitting a discrete distance field. 
        However, solving the system of linear equations associated with this fitting problem is time and memory-consuming, especially in three-dimensional scenarios. 
        Although the iterative method, such as the least squares progressive-iterative approximation (LSPIA) algorithm~\cite{deng2014progressive}, is a superior approach for the fitting problem of B-splines, ensuring the periodicity of the B-splines throughout the iteration process is challenging. 
        Consequently, this subsection proposes a constrained iterative fitting algorithm to address this issue.
    
    Given grid center positions $\{(u_s,v_s,w_s)\}_{s=0}^{S-1}$ and their corresponding values $\{X_s\}_{s=0}^{S-1}$, the fitting problem is formulated as a constrained least-square fitting problem:
    \begin{equation}
        \begin{aligned}
        \label{eq:fitting optimization}
        \min \quad & \sum_{s=0}^{S-1}\|\phi(u_s,v_s,w_s)-X_s\|^2 \\
        \text{s.t.} \quad & \phi(u,v,w)= \sum_{i=0}^{n_u-1}\sum_{j=0}^{n_v-1}\sum_{k=0}^{n_w-1}R_{ijk}(u,v,w)C_{ijk}  \\
        & C_{ijk} = C_{n_u-1-i,j,k}, \quad 0\leq i < r_u  \\
        & C_{ijk} = C_{i,n_v-1-j,k}, \quad 0\leq j < r_v  \\
        & C_{ijk} = C_{i,j,n_w-1-k}, \quad 0\leq k < r_w,   
        \end{aligned}
    \end{equation}
    where $\phi(u,v,w)$ is a periodic B-spline with a symmetric degree of $\mathbf{r}=(r_u,r_v,r_w)$. 
    
    Considering the symmetry of the control coefficients, the basis functions with the same control coefficients can be integrated into a new basis.
    \begin{equation}
    {\hat{N}_{i,p_u}(u)} =
    \begin{cases}
        {N_{i,p_u}(u)},\quad r_u\leq i \leq n_u-r_u-1 \\
        {N_{i,p_u}(u)+N_{n_u-i-1,p_u}(u)}, \quad 0\leq i\leq r_u-1,
    \end{cases}
    \end{equation}
    $\hat{N}_{j,p_v}(v)$ and $\hat{N}_{k,p_w}(w)$ are defined in the same way.
    
    $\phi(u,v,w)$ can be written in the following form:
    \begin{equation}
    \phi(u,v,w) =\sum_{i=0}^{l_u-1}\sum_{j=0}^{l_v-1}\sum_{k=0}^{l_w-1}\alpha_{ijk}(u,v,w)C_{ijk},
    \end{equation}
    where $\alpha_{ijk}(u,v,w)=\hat{N}_{i,p_u}(u)\hat{N}_{j,p_v}(v)\hat{N}_{j,p_v}(v)$, $l_u =n_u-r_u-1 $, and the same applies to $l_v$ and $l_w$.
    
    In the first step of iterations, the initial spline function is represented by:
    \begin{equation}
        \label{eq:initial periodic B-spline}
        \phi^{(0)}(u,v,w)=\sum_{i=0}^{l_u-1}\sum_{j=0}^{l_v-1}\sum_{k=0}^{l_w-1}\alpha_{ijk}(u,v,w)C^{(0)}_{ijk}.
    \end{equation}
    
    All the $l_u\times l_v\times l_w$ control coefficients in Equation~\ref{eq:initial periodic B-spline} are optimized. For each basis function $\alpha_{ijk}$, the data points are assigned to the corresponding set:
    \begin{equation}
        I_{ijk}=\{s:\thinspace\alpha_{ijk}(u_s,v_s,w_s)\neq 0\}.
    \end{equation}
    
    Assuming $(q-1)$ iterations have been performed, the spline function is:
    \begin{equation}
         \phi^{(q-1)}(u,v,w)=\sum_{i=0}^{l_u-1}\sum_{j=0}^{l_v-1}\sum_{k=0}^{l_w-1}\alpha_{ijk}(u,v,w)C^{(q-1)}_{ijk}.
    \end{equation}
    
    The difference for data point $X_s$ is:
    \begin{equation}
        \delta^{(q)}_s = X_s-\phi^{(q-1)}(u_s,v_s,w_s).
    \end{equation}
    
    The difference for control coefficient $C_{ijk}$ is:
    \begin{equation}
        \Delta^{(q)}_{ijk}=\frac{\sum_{s\in I_{ijk}}\alpha_{ijk}(u_s,v_s,w_s)\delta_s^{(q)}}{\sum_{s\in I_{ijk}}\alpha_{ijk}(u_s,v_s,w_s)}.
    \end{equation}
    
    The new control coefficients are obtained by adding the difference to the old control coefficients:
    \begin{equation}
        C_{ijk}^{(q)}=C_{ijk}^{(q-1)}+\Delta_{ijk}^{(q)}.
    \end{equation}
    
    Then the $q$-th spline function is obtained:
    \begin{equation}
        \label{eq:q-th spline function}
      \phi^{(q)}(u,v,w)=\sum_{i=0}^{l_u-1}\sum_{j=0}^{l_v-1}\sum_{k=0}^{l_w-1}\alpha_{ijk}(u,v,w)C^{(q)}_{ijk}.   
    \end{equation}
    
    Despite calculating only $l_u\times l_v\times l_w$ control coefficients, the symmetry described in Definition~\ref{def:symmetric degree} allows for the recovery of the remaining $(n_u-l_u)\times (n_v-l_v)\times (n_w-l_w)$ control coefficients using Equation~\ref{eq:q-th spline function}, resulting in the $q$-th periodic B-spline as follows:
    \begin{equation}
        \phi^{(q)}(u,v,w) = \sum_{i=0}^{n_u-1}\sum_{j=0}^{n_v-1}\sum_{k=0}^{n_w-1}R_{ijk}(u,v,w)C_{ijk}^{(q)}
    \end{equation}
    
    The computational complexity of this iterative algorithm is $O(N \times S)$, where $N$ is the number of control coefficients and $S$ is the number of data points. 
        As proven in the Supplementary Material, this algorithm converges to the constrained least squares fitting problem. 
        After the iteration stops, a periodic B-spline function that represents a porous unit is obtained.
    
    \section{Design and optimization}
    \label{sec: Discussion}
    In Section~\ref{sec: Method}, we propose a novel representation method and corresponding design technique for implicit units.  
    Given a porous sample represented by voxels, the discrete distance field is calculated using a DTM function. 
        Subsequently, the discrete field is fitted using the CON-LSPIA algorithm to obtain a periodic B-spline function that represents an implicit unit. 
        In this section, we firstly discuss the methods for obtaining the porous sample using forward or reverse design. 
        Secondly, there are cases where the fitted periodic B-spline, which represents an implicit porous unit, may be disconnected. 
        In such cases, we discuss methods to optimize the connectivity of the periodic B-spline while preserving its symmetric degree $r$ to maintain the periodicity or symmetry of the porous units. 
        Finally, we employ the implicit unit in minimum compliance problems. 
        The aim of this section is to comprehensively explain the entire process (from design to application) of the implicit units based on the proposed representation method. 
        This will effectively demonstrate the potential applications and advantages of the proposed representation method.
        
    \subsection{Design of implicit units}
    \label{subsec: Design of implicit units}
\begin{figure} [htbp]
        \centering
        \subfloat[]{
            \includegraphics[width=0.18\textwidth]{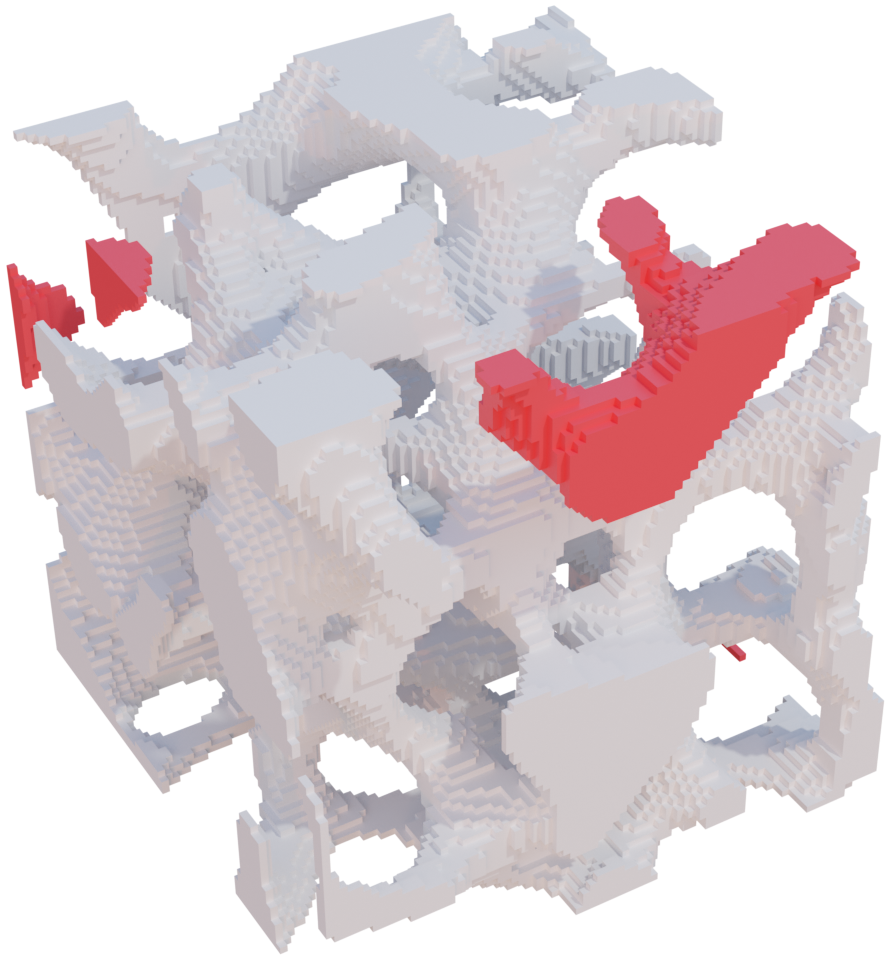}}
        \subfloat[]{
            \includegraphics[width=0.18\textwidth]{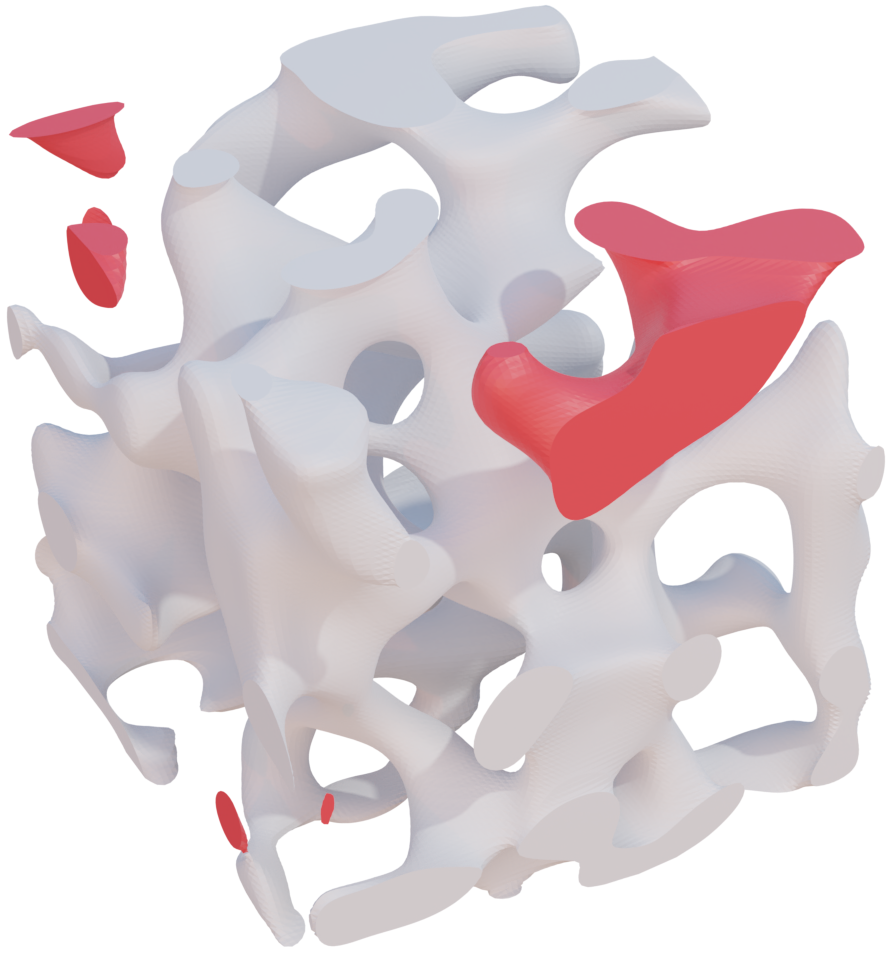}} 
        \subfloat[]{
            \includegraphics[width=0.20\textwidth]{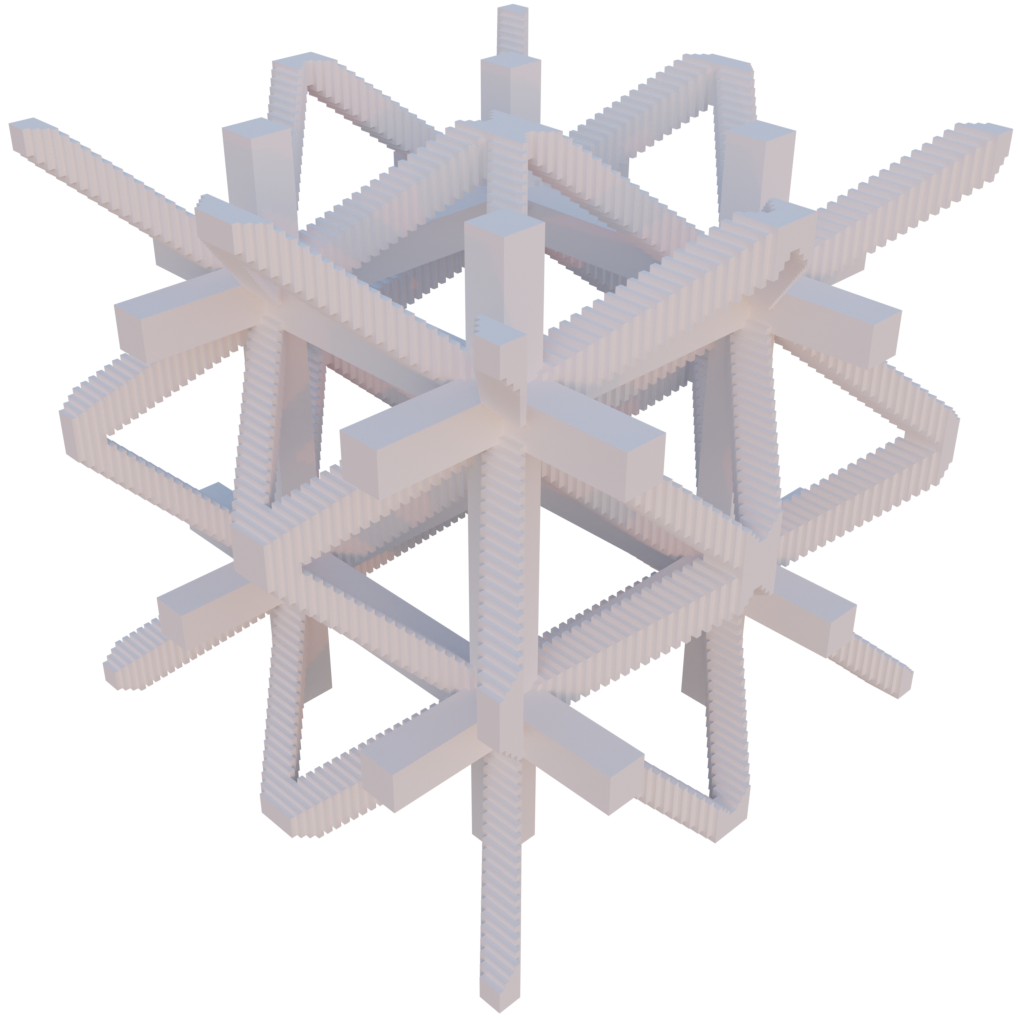}}
        \subfloat[]{
            \includegraphics[width=0.20\textwidth]{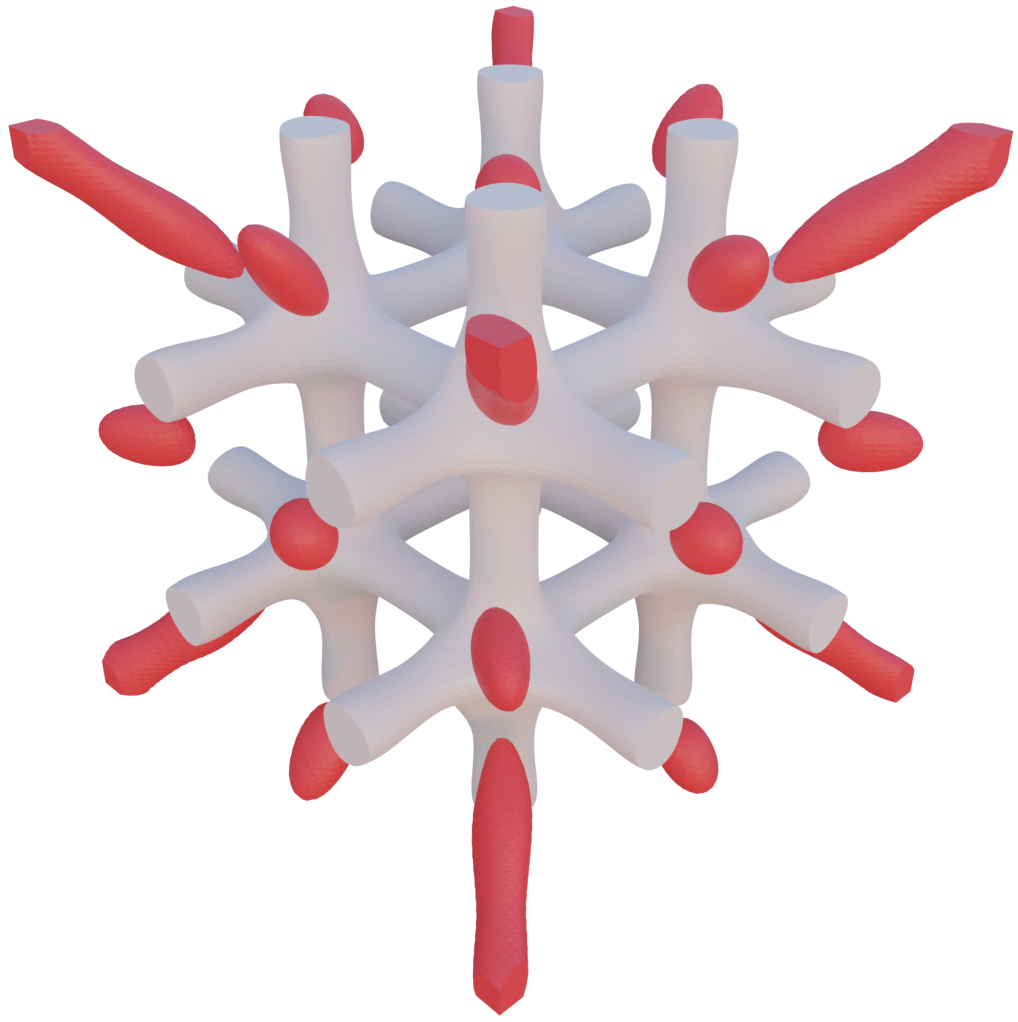}}
        \caption{Implicit units obtained from artificially designed voxels and natural samples. Additional connected components are highlighted in red. (a) A voxel representation of a natural porous sample. (b) A periodic B-spline representation of a periodic porous unit. (c) A voxel representation of an artificially designed sample. (d) A periodic B-spline representation of a symmetric porous unit. }
        \label{fig:Design of implicit units} 
    \end{figure}

    As porous samples represented by voxels can be converted into implicit units using our method, the design of implicit units can be achieved through the design of porous samples. 
        Additionally, geometric structures represented by voxels can be regarded as binary images. 
        From the perspective of reverse design, natural 3D images can be obtained as porous samples using techniques such as CT. 
        {Furthermore, 3D images can be reconstructed from high-resolution 2D images using deep learning methods or statistical methods~\cite{zhang2021slice,ding2018improved}.} 
        As shown in Figure~\ref{fig:Design of implicit units}(a), we select a natural structure as the porous sample. 
        The discrete distance field of the sample in Figure~\ref{fig:Design of implicit units}(a) is fitted with a periodic B-spline with a symmetric degree of 1 in each direction. 
        Figure~\ref{fig:Design of implicit units}(b) exhibits the porous unit with the same relative density as Figure~\ref{fig:Design of implicit units}(a) under the representation of the periodic B-spline.
    
    From the perspective of forward design, the union and intersection operations between voxels can be equivalently viewed as the "OR" and "AND" operations between binary values of $0$ and $1$, significantly enhancing the computational speed of intersection calculations. 
        Therefore, it is possible to interactively design porous samples using Constructive Solid Geometry (CSG). 
        Subsequently, implicit porous units can be quickly obtained using the proposed method, solving the problem of difficult interactive design for implicit porous units. 
        We utilize several simple bars, which are represented by voxels, to design a porous unit with CSG, as shown in Figure~\ref{fig:Design of implicit units}(c). 
        A periodic B-spline with a symmetric degree of $\textbf{r}=\lfloor {\frac{n}{2}} \rfloor$ in each direction is used to fit the discrete distance field and obtain a symmetric unit as shown in Figure~\ref{fig:Design of implicit units}(d). 
    
    The fitted porous units are smooth and can be spliced infinitely, making the design method under the new representation effective in generating implicit porous units through forward and inverse design. 
        Although this work assumes voxel representation for porous samples, as discussed in Section~\ref{subsec:Construction of discrete distance field}, B-rep and F-rep representations are also applicable in this study. 
        We select voxel representation as the starting point of the design due to its efficient intersection calculation between voxels. 
        Complex voxel units can be interactively designed through Constructive Solid Geometry. 
        Additionally, existing CT technology provides abundant 3D binary images of natural porous structures, which can be directly utilized to design implicit units. 
        Therefore, this study demonstrates potential in the field of biomimetics. 
        
    \subsection{Connectivity Optimization}
    \label{subsec: Optimization of topological features}
    In the previous section, we demonstrate that by combining the forward and inverse design of voxels with the proposed method, it is possible to design implicit porous units. 
        However, as shown in Figure~\ref{fig:Design of implicit units}, when converting voxels into implicit units, many additional connected components are introduced. 
        This is because the porous sample itself is not connected, and on the other hand, symmetry is imposed on asymmetric porous samples. 
        To address this issue, we propose an optimization method for ensuring connectivity of the implicit units in this subsection. 
        
    Given a trivariate B-spline function $\phi:~\mathbb{R}^3 \rightarrow \mathbb{R}$ and a constant $t$, a set $A_t$ can be defined as follow: 
    \begin{equation}
        A_t = \{(u,v,w)~|~\phi(x,y,z) \leq t\}.
    \end{equation}
        The set $A_t$ can be discretized to compute its zero-dimensional topological features, specifically, the connected components, using persistent homology~\cite{edelsbrunner2002topological,chazal2021introduction}. 
        Furthermore, for a varying value $t$, persistent homology can capture the changes in 0-dimensional topological features of set $A_t$ as $t$ gradually increases from $-\infty$ to $+\infty$. 
        Assuming a connected component appears in set $A_{b}$ at $t=b$, and merges with other components in the set $A_{d}$ at $t=d$, it can be represented as a 0-dimensional persistence pair $(b, d)$. 
        Persistent homology can record all the 0-dimensional persistence pairs that appear and disappear during the increase of $t$. 
        For a more rigorous definition based on algebraic topology, we recommend referring to the work by Edelsbrunner et al~\cite{edelsbrunner2022computational}. 
    
    Assuming that all 0-dimensional persistent pairs $\{(b_i,d_i)\}_{i=0}^{\Lambda-1}$ corresponding to the periodic B-spline function $\phi$ have been sorted in descending order based on their death times. 
        Since the set $A_{+\infty} = [0,1]\times[0,1]\times[0,1]$ at $t=+\infty$ is a unit cube, there exists only one connected component, which means $d_0=\infty$. 
        As described in section~\ref{subsec:Construction of discrete distance field of porous model}, an implicit porous is represented by $\phi(u,v,w)\leq c(u,v,w)$. 
        The death time $d_1$ is minimized to eliminate the additional connected components in the implicit porous. 
        Therefore, when $c(u,v,w)$ is greater than $d_1$, there will be only one connected component. 
        To eliminate the additional connected components of the implicit units, we formulate the following optimization problem: 
        \begin{equation}
            \begin{aligned}
        \label{eq:topological features optimization}
        \min \quad & L=d_1 \\
        \text{s.t.} \quad & \text{The symmetric degree of $\phi$ is fixed}. 
            \end{aligned}
        \end{equation}
    
    Similarly to previous studies in the literature~\cite{dong2022topology,bruel2020topology,poulenard2018topological}, a persistent inverse mapping function, denoted as $\pi_\phi$, is employed to establish a correspondence between persistent pairs and locations in the parameter domain: 
    \begin{equation}
        ((u_b,v_b,w_b),(u_d,v_d,w_d)) = \pi_\phi(b,d).
    \end{equation}
    
    The derivative of $L$, with respect to the control coefficient $C_{ijk}$ of $\phi(u,v,w)$ can be computed as: 
    \begin{equation}
            \frac{\partial L}{\partial C_{ijk}} = \frac{\partial d_1^{(0)}}{\partial C_{ijk}} 
            = \alpha_{ijk}(\pi_\phi(d_1)). 
    \end{equation}
    
    Subsequently, the value of $L$ is reduced using gradient descent. 
        {In our experiment, the termination condition for iteration is $L<0$.
        This means that when $c(u,v,w) \geq 0$ is reached, the optimized porous model $\phi(u,v,w)\leq c(u,v,w)$ will have only one connected component.}
        It is important to note that, similar to the approach described in Section~\ref{subsec: fitting}, we do not directly optimize all $n_u \times n_v \times n_w$ control coefficients of the periodic B-spline function $\phi(u,v,w)$. 
        Due to symmetry, we only optimize $l_u\times l_v \times l_w$ control coefficients. 
        Therefore, during the optimization process, the symmetry of the periodic B-spline is ensured, the constrained optimization problem in Equation~\ref{eq:topological features optimization} is transformed into an unconstrained optimization problem.
    
    \subsection{Topology optimization}
    \label{subsec: topology optimization}
    In this subsection, we apply implicit porous units to minimize compliance of a porous model. 
        Compared to the non-periodic implicit porous structure represented by B-spline functions in work~\cite{gao2022connectivity}, due to the periodicity of the porous unit, we do not need to transform the entire porous structure into mesh for topology optimization. 
        We treat the porous unit as a new material using homogenization methods. 
        Therefore, the computational burden and memory usage in topology optimization can be significantly reduced. 
      
    We treat the porous unit as a new material. 
        The homogenized elastic tensor of this new material can be determined through homogenization methods~\cite{dong2019149}. 
        For a porous unit, the homogenized elastic tensor $C^H$ takes the form of a symmetric matrix $C^H=[c_{ij}]_{6\times 6}$ that meets $c_{ij}=c_{ji}$. 
    \begin{figure} [htbp]
     \vspace{-0.4cm}
        \centering
        \subfloat[]{
            \includegraphics[width=0.20\textwidth]{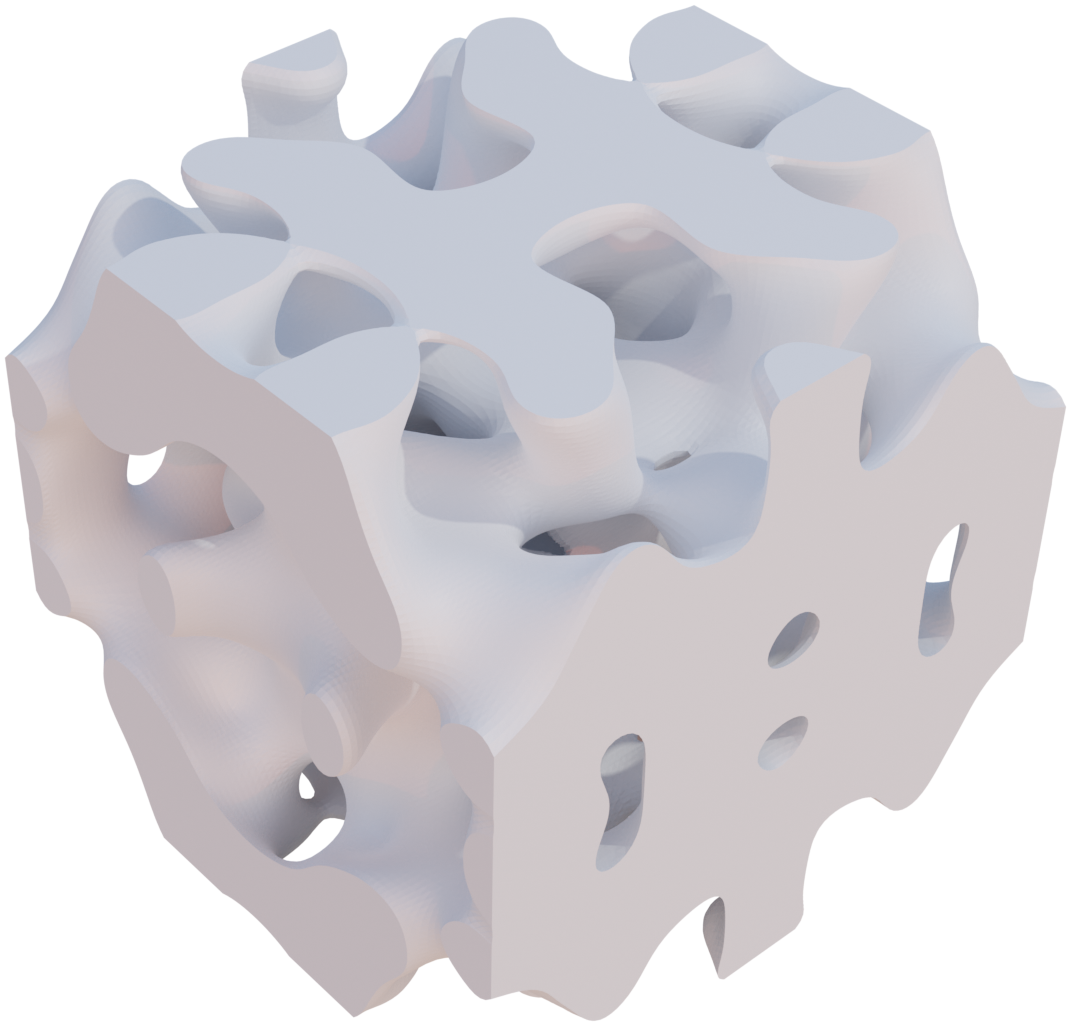}} 
        \subfloat[]{
            \includegraphics[width=0.5\textwidth]{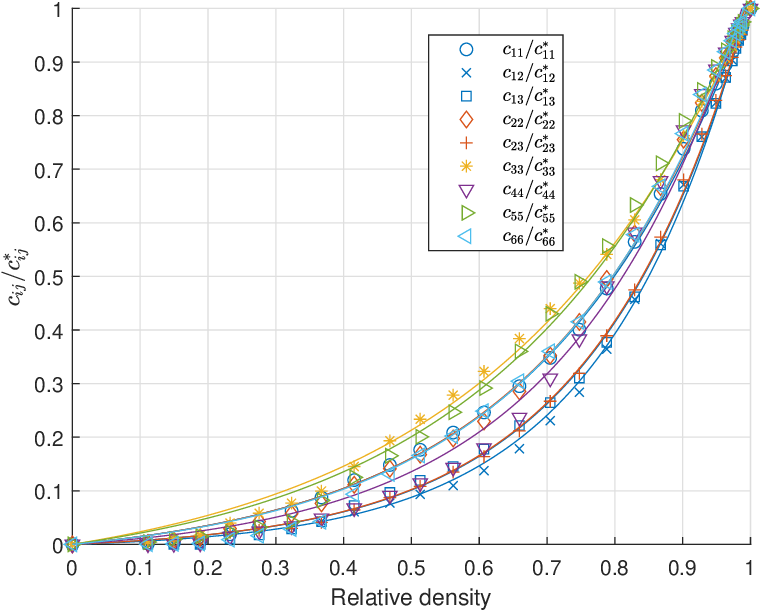}}
        \caption{Relationship between density and homogenized elastic tensor. (a) A symmetric porous unit. (b) The fitted curves. }
        \label{fig:topology optimization}
    \end{figure}
    
    The "Gibson-Ashby" model suggests that the relative elastic tensor of a porous structure is a function of its relative density, which has been widely used in topology optimization of porous structures~\cite{li2019design,feng2022stiffness}. 
        Considering a symmetric porous unit represented by a periodic B-spline as shown in Figure~\ref{fig:topology optimization}(a), the elastic tensor of the base material is denoted as $C^*=[c_{ij}^*]_{6\times 6}$. 
        Due to its symmetry, the homogenized elastic tensor has $9$ coefficients of freedom. 
        By calculating the homogenized elastic tensor at different relative densities, the relationship depicted in Figure~\ref{fig:topology optimization}(b) can be fitted using the following formula:
    \begin{equation}
        c_{ij}(\rho)/c_{ij}^* = a_1e^{a_2\rho}-a_1,
    \end{equation}
    where $\rho$ is the relative density of the porous unit. The coefficients $a_1$ and $a_2$ are parameters that need to be fitted. 
    
    The aim of topology optimization is to find an optimal density field for the mechanical stiffness problem. 
        A density field represented by a B-spline function is utilized to ensure the smoothness of the density distribution: 
        \begin{equation*}
            \rho (u,v,w)=\sum_{i=0}^{N_u-1}\sum_{j=0}^{N_v-1}\sum_{k=0}^{N_w-1}R_{ijk}(u,v,w)\rho_{ijk}.
        \end{equation*}
        The mathematical expression of the minimum compliance problem is as follows:
    \begin{equation}
        \begin{aligned}
        \label{eq:topology optimization}
        \min_{\rho_{ijk}} \quad & \mathcal{C}(\rho)=U^T K U \\
        \text{s.t.} \quad & KU=F  \\
                          & K= \sum_{e=1}^{N_e}\int_{\Omega_e}B_e^T C^H(\rho) B_e d\Omega  \\
                          & V = \int_{\Omega}\rho(u,v,w)d\Omega = \vartheta  \text{Vol}(\Omega)  \\
                          & \rho_{min}\leq \rho_{ijk} \leq \rho_{max} ,   
        \end{aligned}
    \end{equation}
    where the compliance $\mathcal{C}(\rho)$ is the objective function, the control coefficients $\rho_{ijk}$ are the optimization variable, $K$ is the global stiffness matrix, $U$ is the displacement, $F$ is the load vector, $C^H$ is the homogenized elastic tensor which is the function of $\rho$, $N_e$ is the number of elements, $V$ is the volume of the porous model, $\Omega$ is the physical domain of the B-spline solid, $\vartheta$ is the fraction of the volume, and Vol$(\Omega)$ is the volume of the B-spline solid. 
        The first and second constraints are the equilibrium equations established through iso-geometric analysis (IGA). 
        The third constraint involves limiting the volume ratio. 
        The final constraint $\rho_{max}$ and $\rho_{min}$ are the maximal and minimum relative density of the porous units. 
    
    The objective of the optimization problem~(\ref{eq:topology optimization}) is to find a density field for the porous model that minimizes its compliance. 
    This topology optimization problem is solved using Optimality Criteria(OC)~\cite{hassani1998review}.
    Notably that the porous model is defined by the set $\{(u,v,w)~|~\phi(u,v,w)\leq c(u,v,w)\}$, where $\phi(u,v,w)$ is the implicit representation of the porous unit and $c(u,v,w)$ is the threshold distribution field. 
   Increasing the threshold value results in an increase in the relative density of the porous model, which indicates a one-to-one relationship between the threshold distribution field $c(u,v,w)$ and the density field $\rho(u,v,w)$.
   Based on this relationship, the algorithm presented in citation~\cite{hu2021heterogeneous} is employed to compute the threshold field $c(u,v,w)$ given the representation of the porous unit ($\phi(u,v,w)$) and the desired density field ($\rho(u,v,w)$). Subsequently, the resulting threshold distribution field $c(u,v,w)$ is used to generate the porous model.
    
    \section{Results and discussions}
    \label{sec: Implementation, results and discussions}
    The design method proposed in this study is implemented using the C++ programming language and tested on a personal computer equipped with an i7-10700 CPU running at 2.90 GHz and 16 GB of RAM. This section provides the details of the experiments to prove the effectiveness of the method.
    
    \subsection{Analysis of symmetric degree}
    \label{Analysis of symmetric degree}
    As discussed in Section~\ref{subsec: Periodic B-spline}, the size of symmetric regions in periodic B-splines increases as the symmetric degree increases. 
        In this subsection, we analyze the effect of the symmetric degree on the fitted periodic B-spline functions. 
    
    \begin{figure} [htbp]
        \centering
        \subfloat[sample]{
            \includegraphics[width=0.20\textwidth]{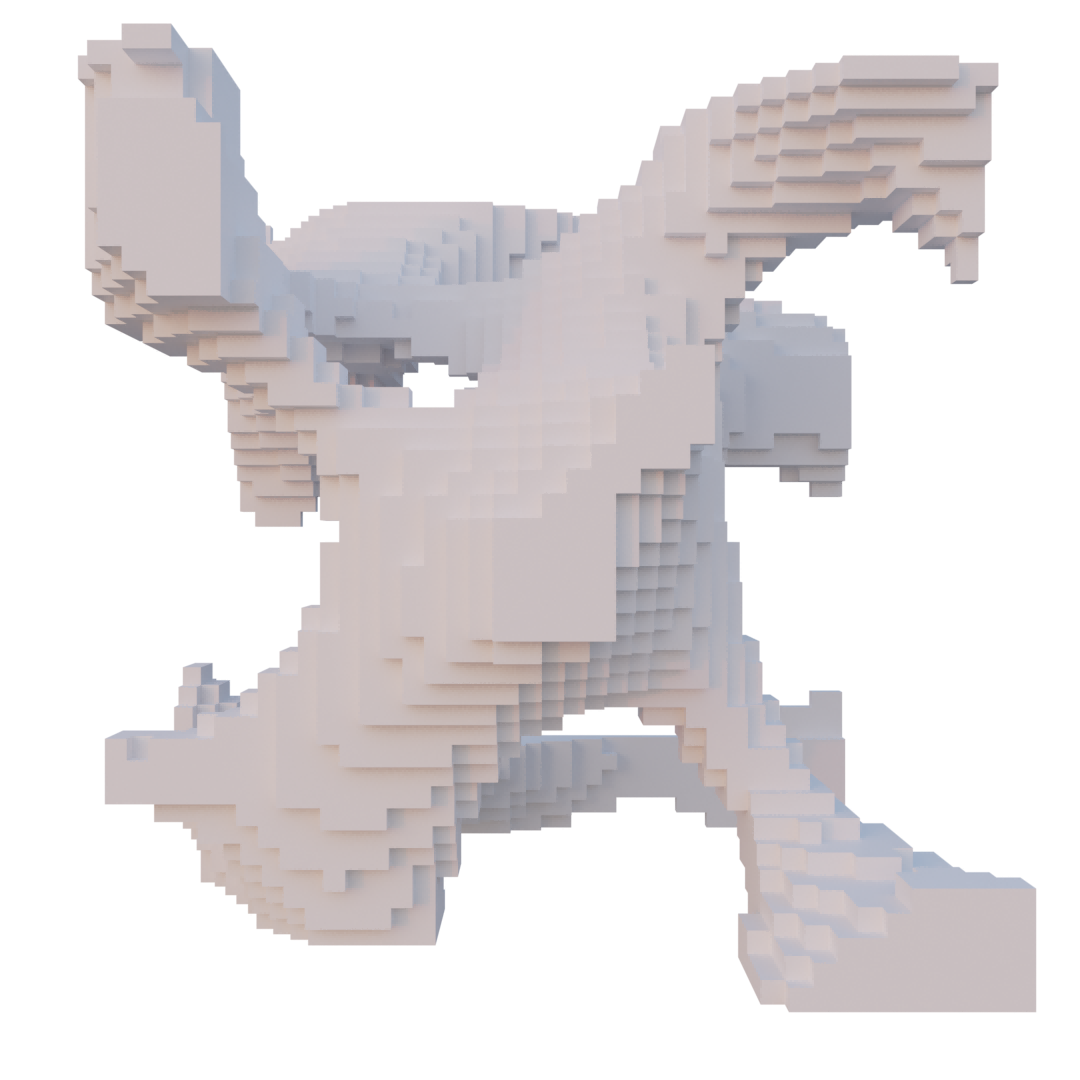}}
        \subfloat[$r=0$]{
            \includegraphics[width=0.2\textwidth]{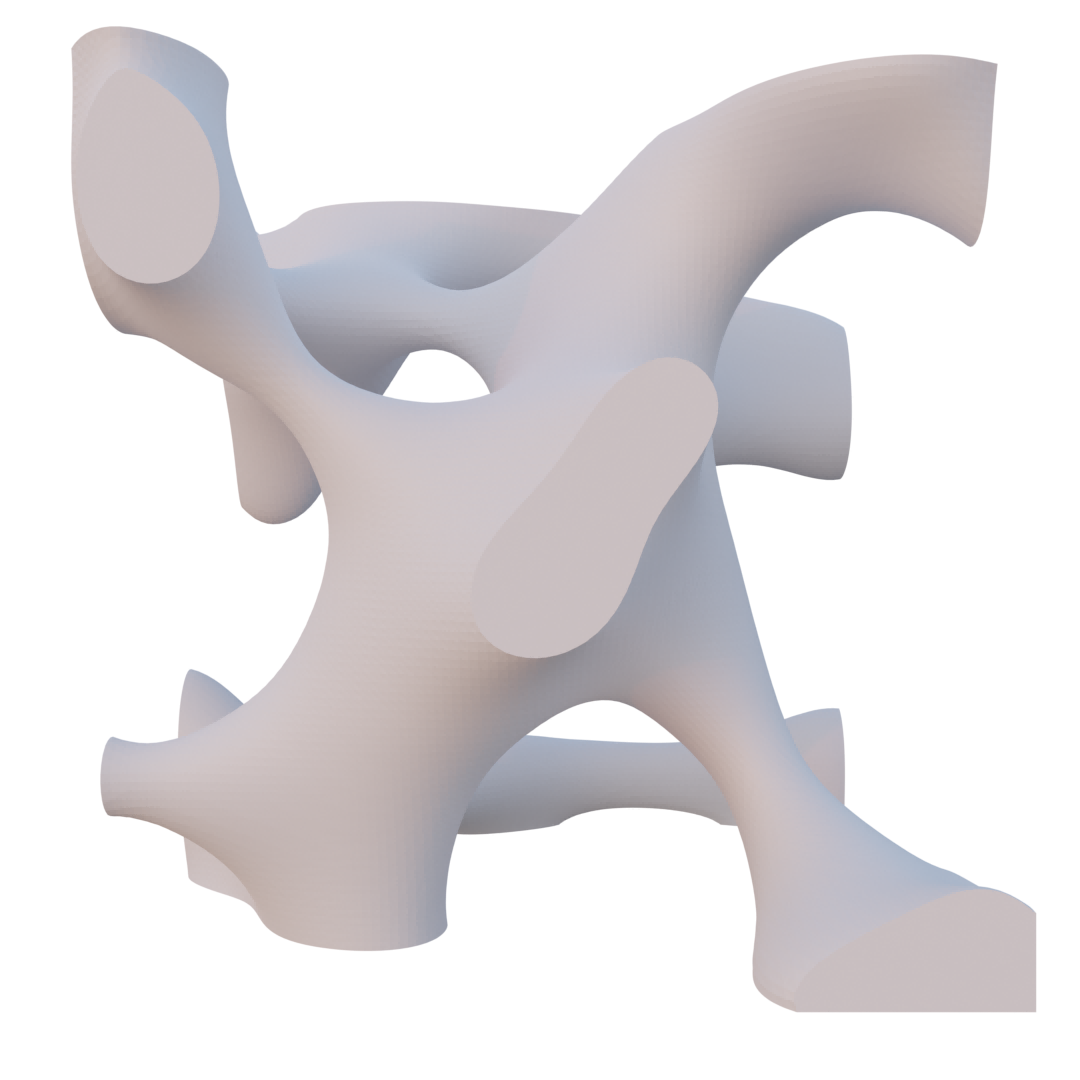}} 
        \subfloat[$r=1$]{
            \includegraphics[width=0.2\textwidth]{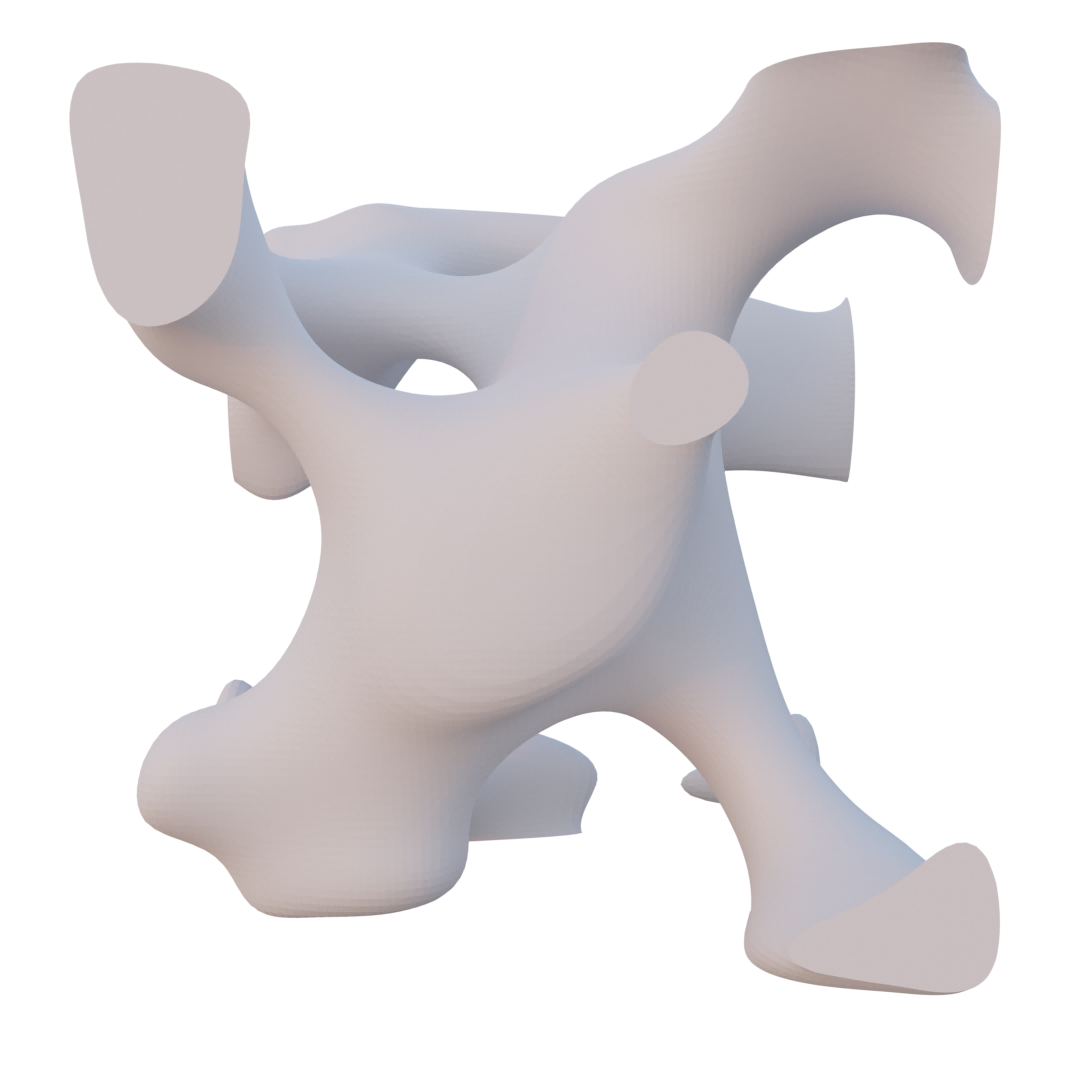}}
        \subfloat[$r=2$]{
            \includegraphics[width=0.2\textwidth]{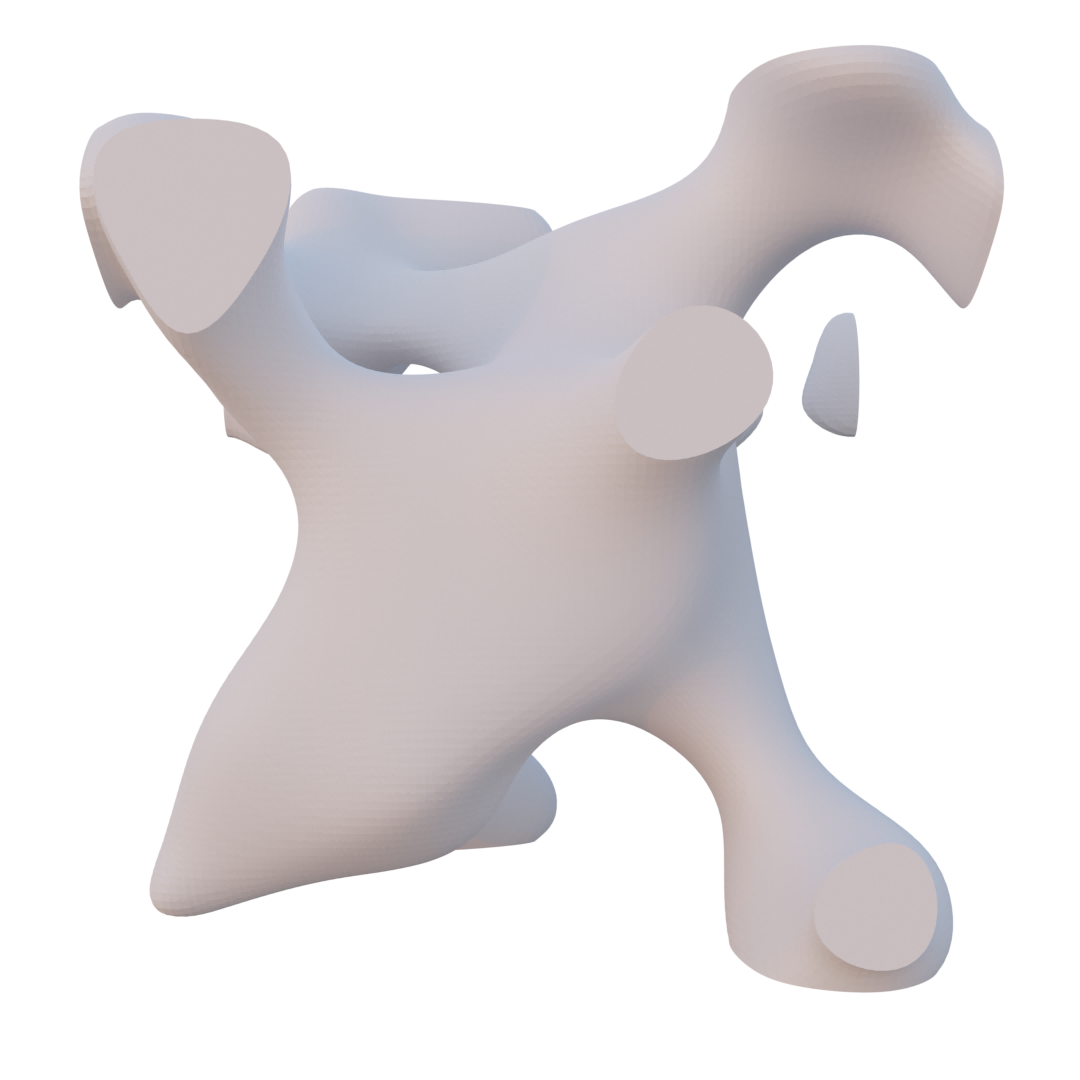}} \\
        \subfloat[$r=3$]{
            \includegraphics[width=0.2\textwidth] {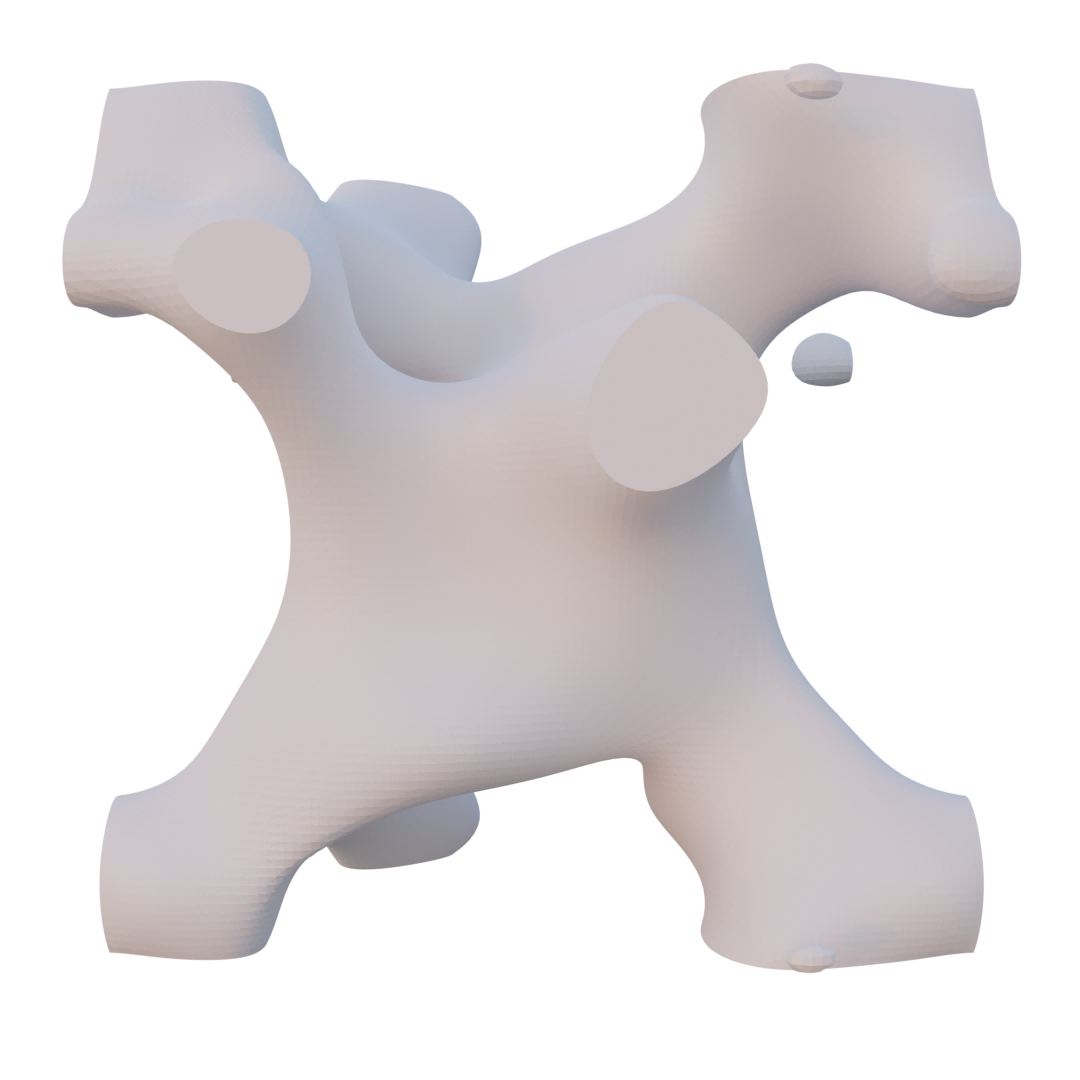}}
        \subfloat[$r=4$]{
            \includegraphics[width=0.2\textwidth]{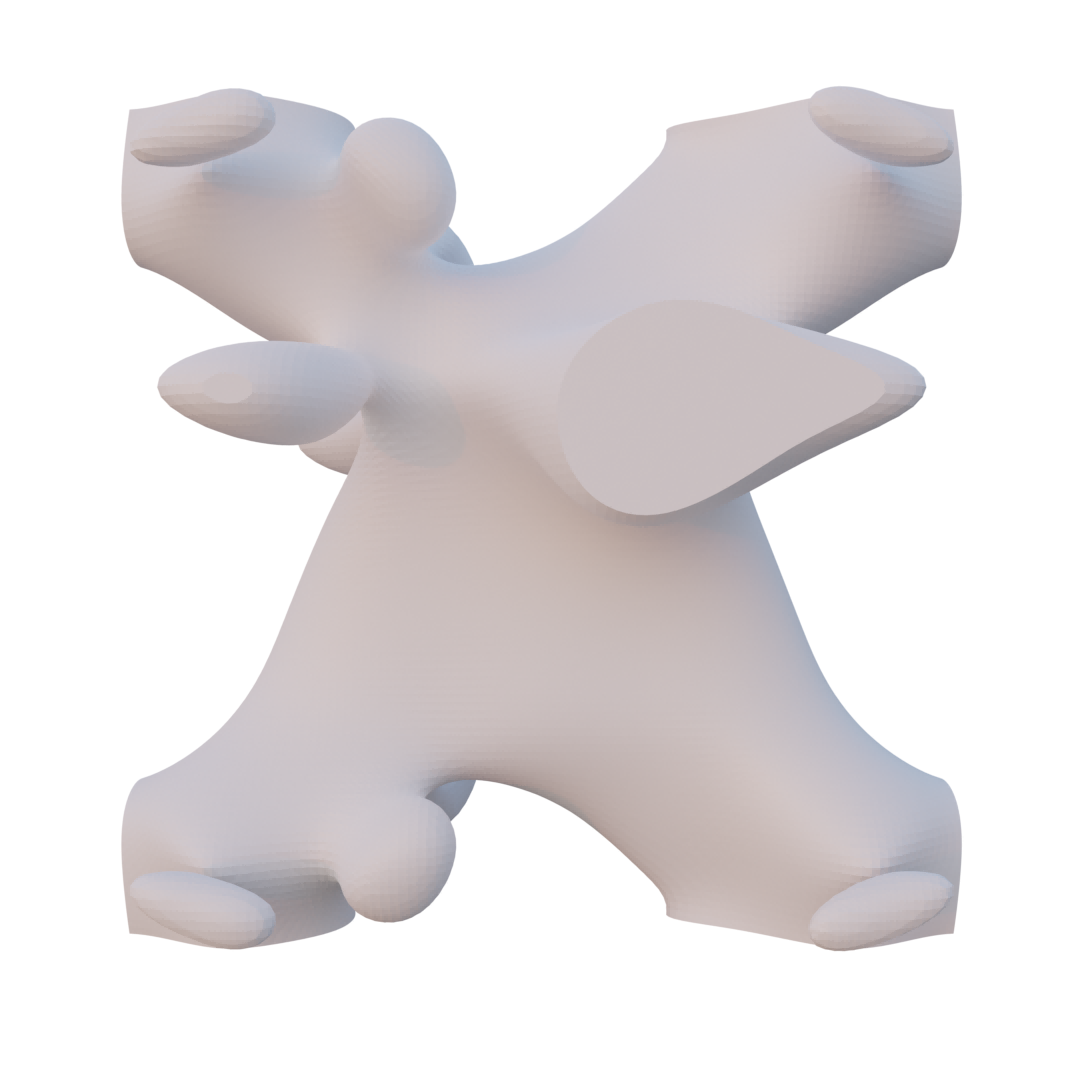}}  
        \subfloat[$r=5$]{
            \includegraphics[width=0.2\textwidth]{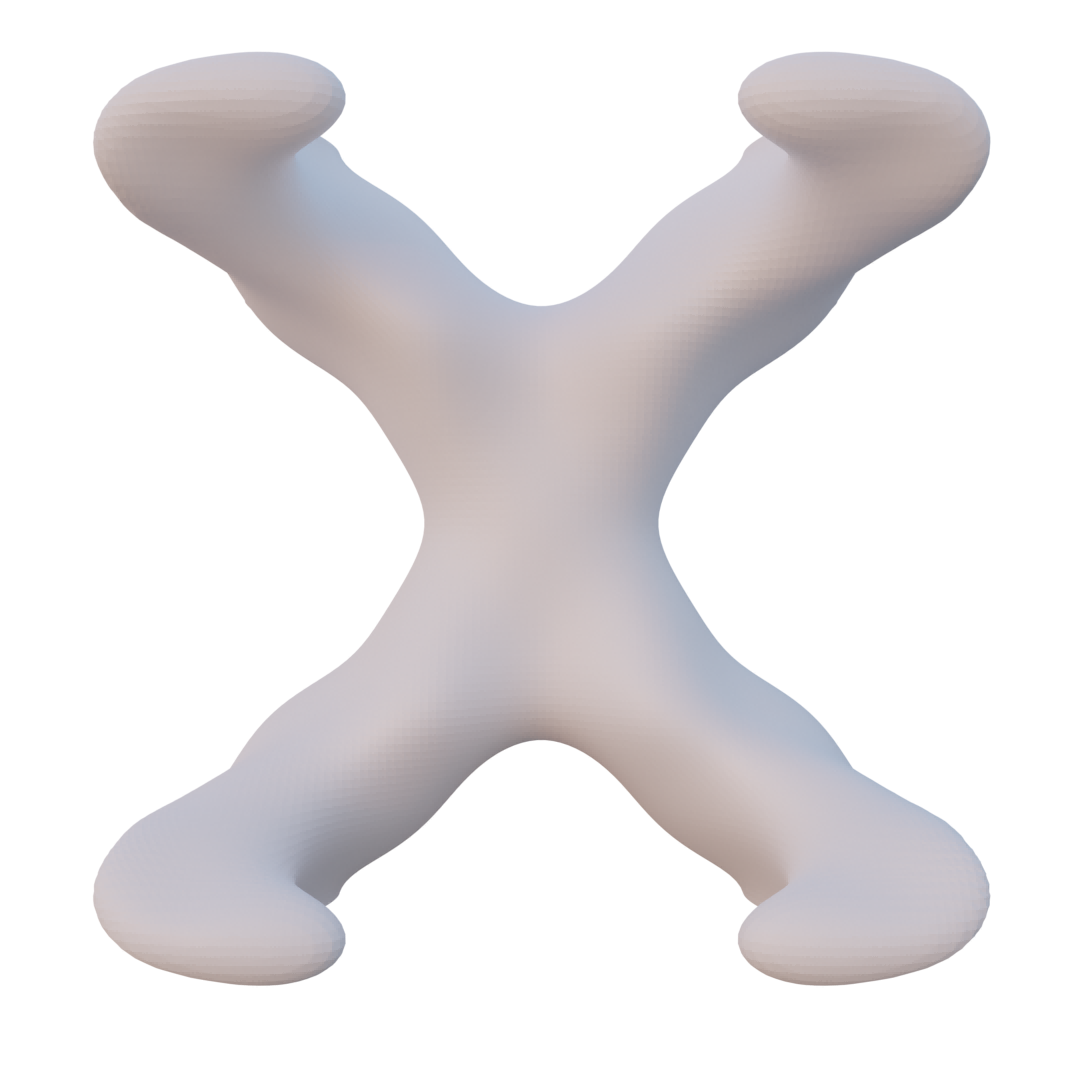}}
        \caption{Fitting results for various symmetric degrees $r$. A porous sample is presented in (a). Figures (b) to (g) display the fitted porous units represented by periodic B-splines with the same density as the porous sample.}
        \label{fig:symmetric degree}
    \end{figure}

    Given a porous sample size of $80 \times 80 \times 80$ (see Figure~\ref{fig:symmetric degree}(a)), periodic B-splines with varying symmetric degree $r$ are used to fit the discrete distance field of this sample using CON-LSPIA. 
        The number of control coefficients of these periodic B-splines is $11\times 11\times 11$, and their degree is $3 \times 3 \times 3$. 
        The results of fitting with different symmetric degrees $r$ (the symmetric degree in each direction is $r$) are illustrated in Figure~\ref{fig:symmetric degree}.
        From a visual perspective, an increase in $r$ enhances the symmetry of the porous unit but decreases its resemblance to the sample. 
        When the symmetric degree $r$ is greater than $0$, the porous unit becomes periodic. 
        For a symmetric degree of $\lfloor {\frac{11}{2}} \rfloor = 5$, the porous unit is symmetric. 
        The results suggest that controlling the symmetric degree $r$ allows for effective regulation of the symmetry in the implicit porous unit. 
        Consequently, the proposed representation method can represent symmetric or periodic porous units. 
        
\begin{figure} [htbp]
        \centering
        \subfloat[sample 1]{
            \includegraphics[width=0.2\textwidth]{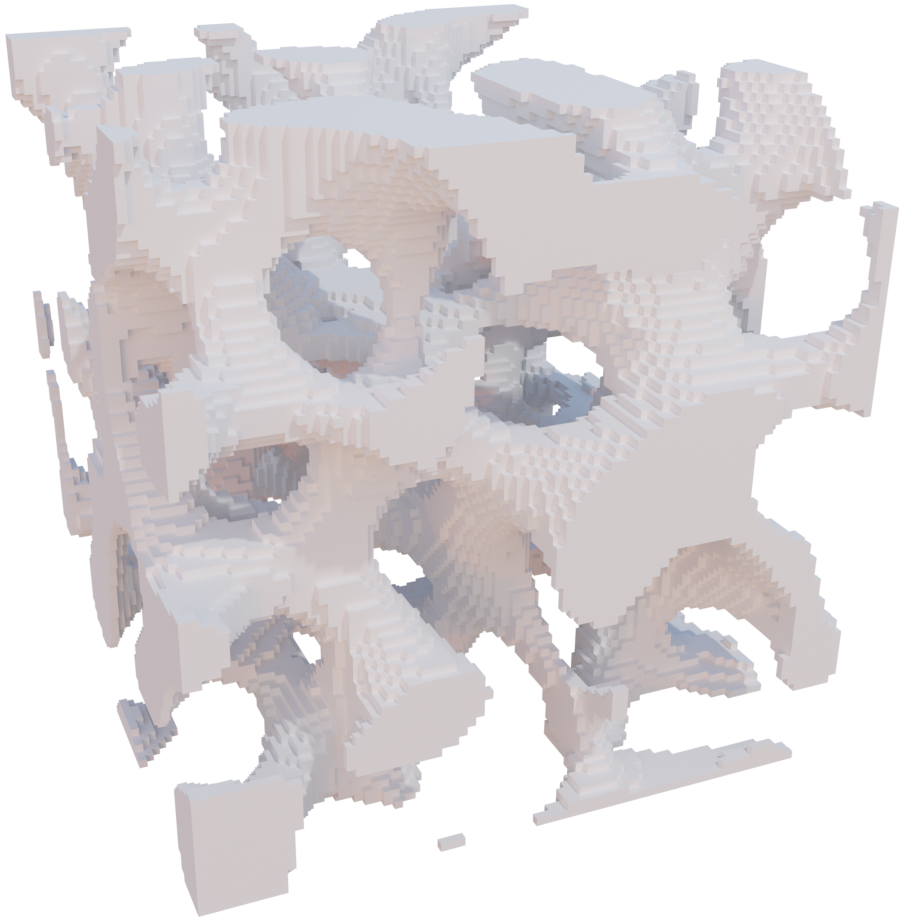}}
        \subfloat[sample 2]{
            \includegraphics[width=0.2\textwidth]{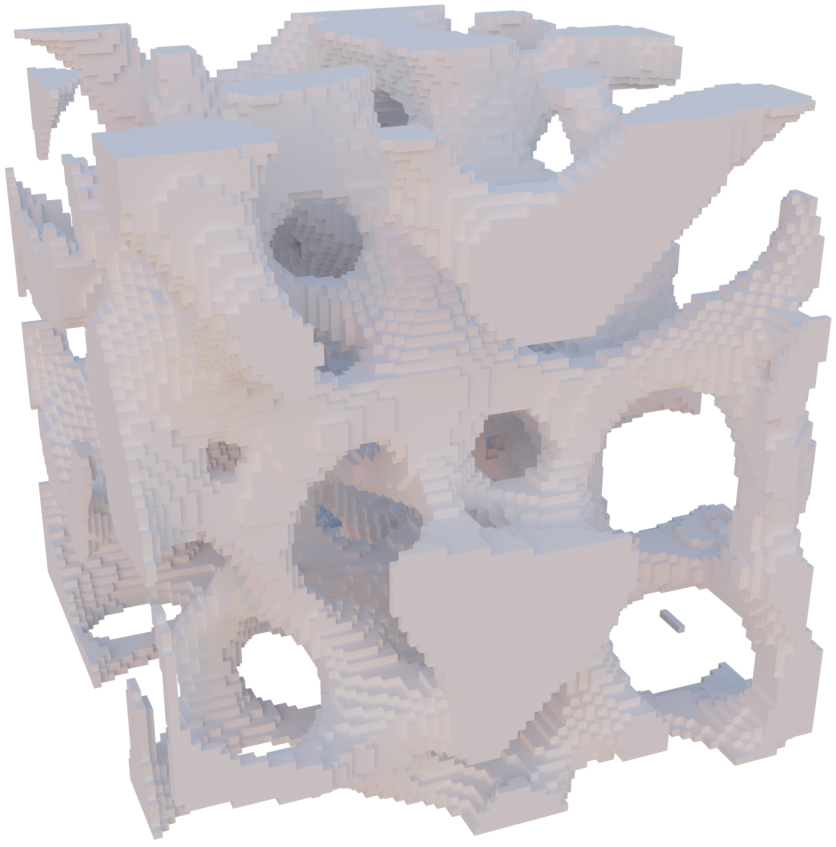}} 
        \subfloat[sample 3]{
            \includegraphics[width=0.2\textwidth]{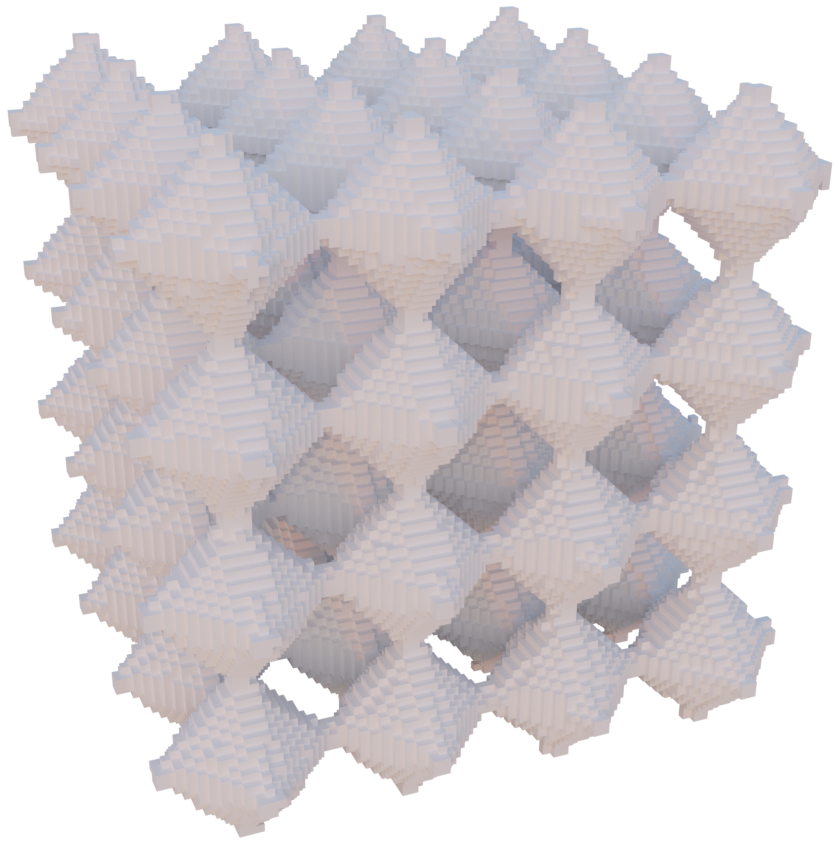}}
        \subfloat[sample 4]{
            \includegraphics[width=0.2\textwidth]{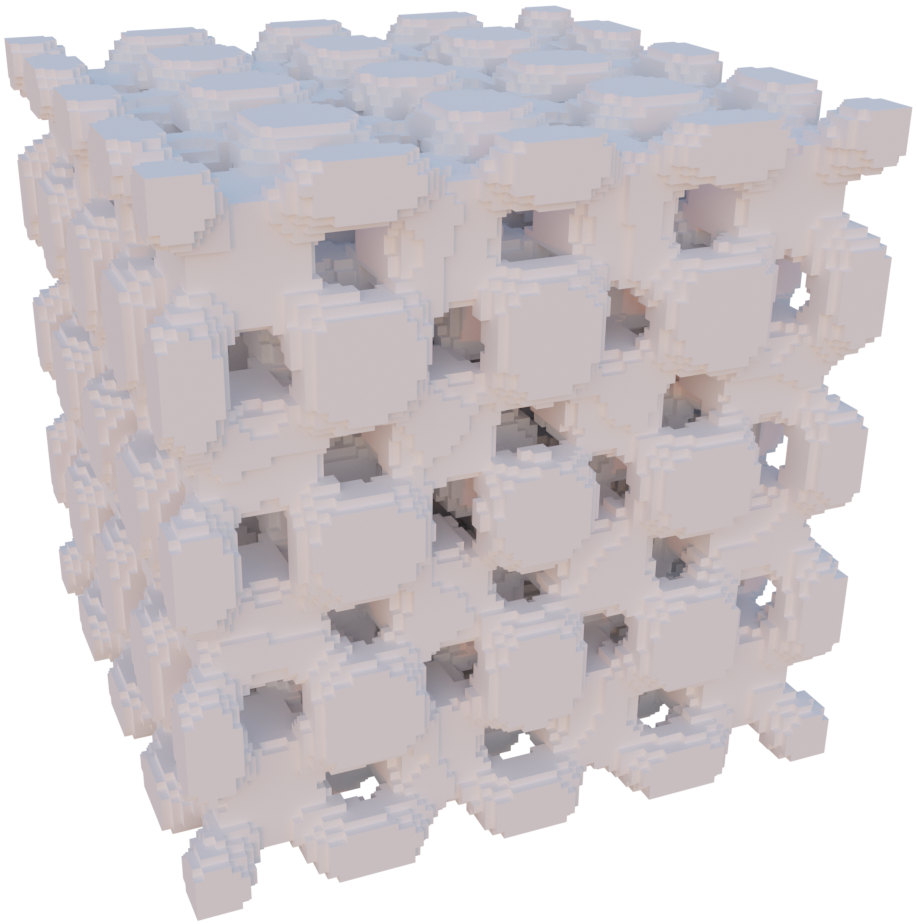}}
        \caption{The porous samples used for experiment. Sample 1 and 2 are non-periodic. Sample 3 and sample 4 are both periodic and symmetric.}
        \label{fig:porous samples} 
    \end{figure}

    To further demonstrate the effectiveness of the CON-LSPIA algorithm, we conduct experiments using four porous samples, as depicted in Figure~\ref{fig:porous samples}, to examine the impact of the symmetric degree $r$ on the fitting precision of the algorithm. 
        Porous samples 1 and 2 are non-periodic, while samples 3 and 4 are both periodic and symmetric. 
        We employ periodic B-splines to fit the discrete distance field of these samples with different symmetric degrees. 
        The Mean Square Errors (MSE) of the fitting processes are recorded in Table~\ref{table: MSE}. 
        For non-periodic porous samples 1 and 2, the MSEs increase as the symmetric degree $r$ increases. 
        For symmetric porous samples 3 and 4, the MSEs exhibit small fluctuations as the symmetric degree increases. 
        To quantify this fluctuation, we calculate the variance of MSEs at different symmetric degrees. 
        The results indicate that, compared to non-periodic samples, the variance of MSEs for symmetric samples is significantly smaller, as the imposition of symmetry alters the inherent structure of non-symmetric samples.
    
    When fitting symmetric samples 3 and 4, due to their inherent symmetry, the solutions obtained through fitting with symmetric constraints and without symmetric constraints should be very close. 
        The small fluctuation in MSEs for samples 3 and 4 at different symmetric degrees supports this deduction, further highlighting the effectiveness of the CON-LSPIA algorithm. 
        The fluctuation in MSEs occurs because, on one hand, an increase in symmetric degree reduces the fitting precision by limiting the freedom of control coefficients. 
        On the other hand, as the symmetric degree increases, the solution obtained by CON-LSPIA aligns with the symmetry of the data being fitted, which increases the fitting precision. 
        Under the influence of both factors, the MSEs exhibit slight fluctuations.
    
    Although imposing symmetric constraints reduces the fitting precision, a visual comparison between Figure~\ref{fig:Design of implicit units}(a) and Figure~\ref{fig:Design of implicit units}(b), as well as between Figure~\ref{fig:symmetric degree}(a) and Figure~\ref{fig:symmetric degree}(c), shows similar structures. 
        Furthermore, the MSEs presented in Table~\ref{table: MSE} demonstrate no significant increase as the symmetric degree increases. 
        Thus, by imposing symmetry without significantly altering the structure, we can obtain periodic porous units.           
        Introducing periodicity provides the advantage of directly applying these units to topology optimization and splicing them into any size, bringing great convenience in the usage.

    \begin{table}[]
    \centering
    \caption{The influence of symmetric degree $r$ on the fitting precision.}
    \label{table: MSE}
    \resizebox{1\textwidth}{!}{
    \begin{tabular}{cccccccllllc}
    \hline
    Sample No. & \multicolumn{10}{c}{MSE under different symmetric degrees $r$}                                                      & Variance               \\ \cline{2-11}
               & 0     & 1     & 2     & 3     & 4     & 5     & 6     & 7     & 8     & 9     &                        \\ \hline
    1       & 0.092 & 0.534 & 1.051 & 1.936 & 3.013 & 3.978 & 4.901 & 5.598 & 6.066 & 6.394 & 5.044                  \\
    2       & 0.101 & 0.548 & 1.111 & 1.942 & 3.017 & 3.981 & 4.802 & 5.381 & 5.868 & 6.245 & 4.684                  \\
    3        & 0.176 & 0.176 & 0.178 & 0.178 & 0.179 & 0.181 & 0.181 & 0.180 & 0.181 & 0.180 & $3.4 \times 10^{-6}$   \\
    4      & 0.265 & 0.266 & 0.270 & 0.273 & 0.276 & 0.265 & 0.264 & 0.270 & 0.272 & 0.272 & $15.01 \times 10^{-6}$ \\ \hline
    \end{tabular}}
    \end{table}
    
    \subsection{Connectivity optimization}
    \label{Connectivity optimization}
    In this subsection, we present examples to demonstrate the effectiveness of the method proposed in Section~\ref{subsec: Optimization of topological features} in eliminating additional connected components. 
        Additionally, we demonstrate that the symmetric degree of the periodic B-spline is maintained during connectivity optimization.
    
\begin{figure} [htbp]
        \centering
            \subfloat[]{
            \includegraphics[width=0.2\textwidth]{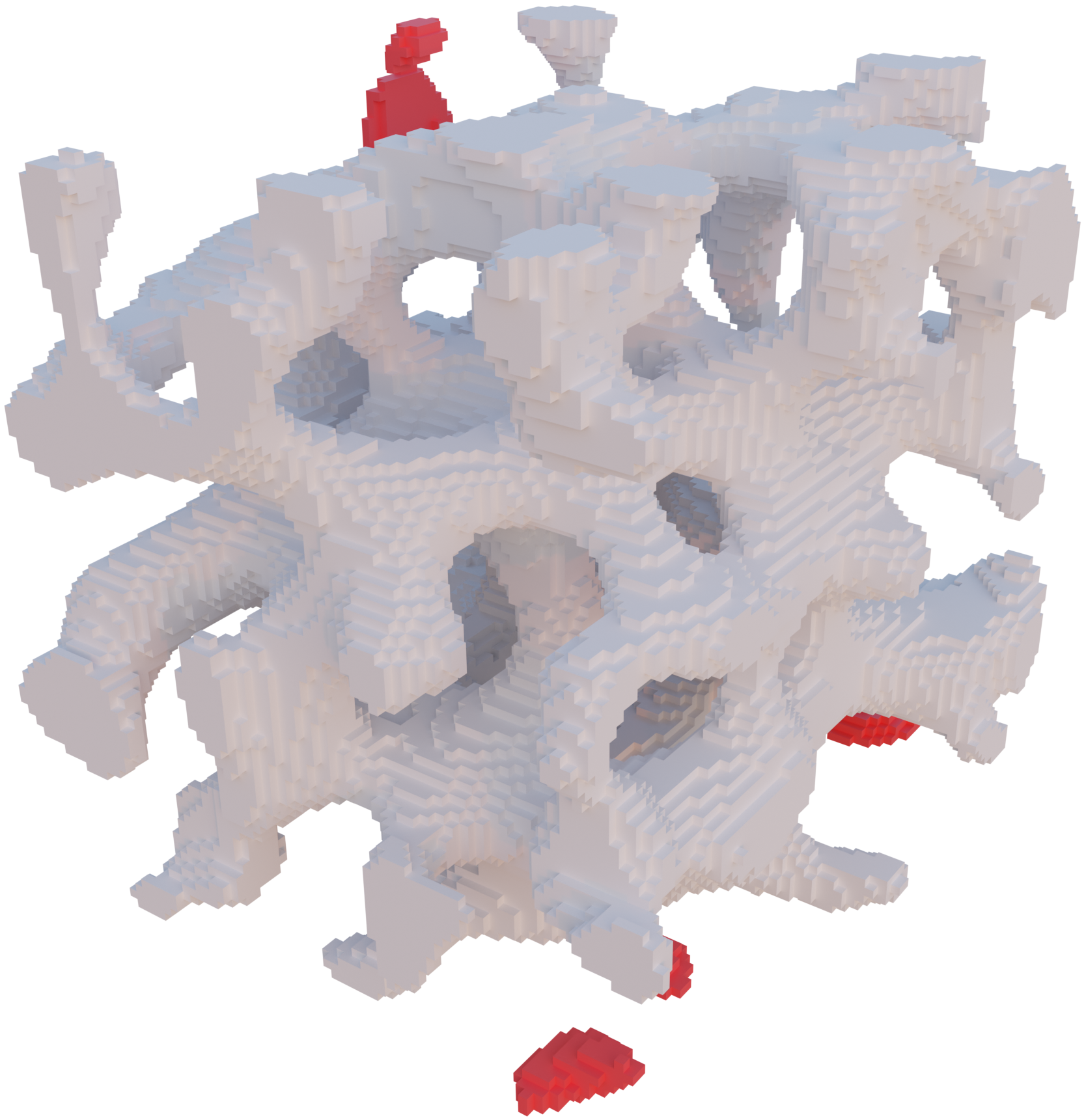}} 
        \subfloat[]{
            \includegraphics[width=0.2\textwidth]{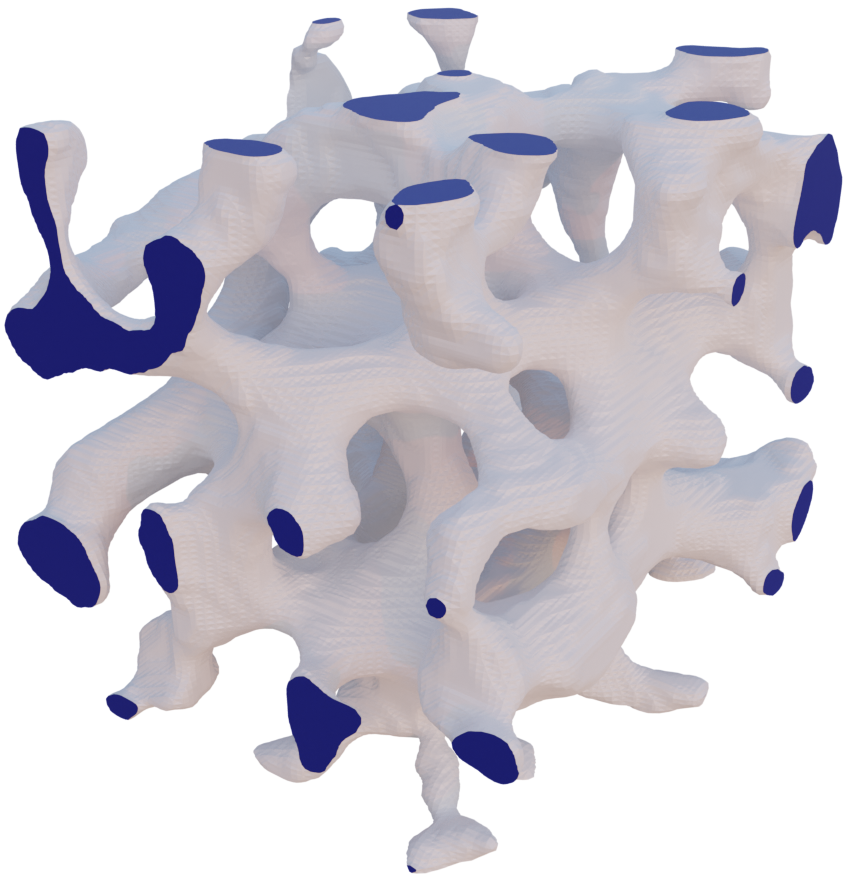}} 
        \subfloat[]{
            \includegraphics[width=0.2\textwidth]{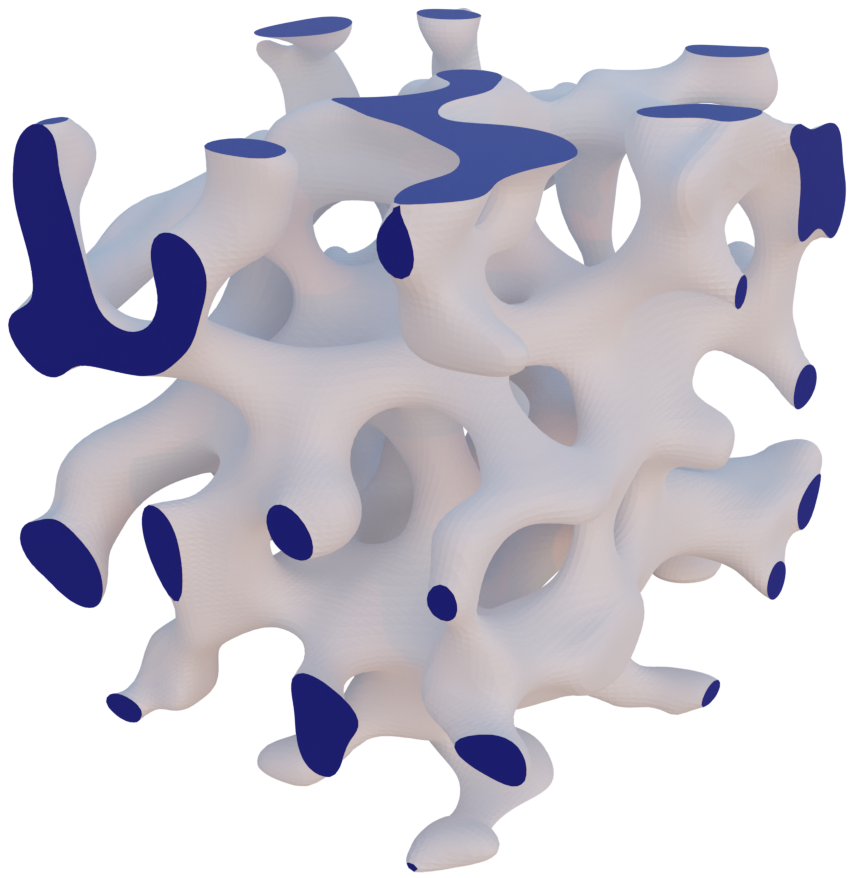}} \\
            \subfloat[]{
            \includegraphics[width=0.35\textwidth]{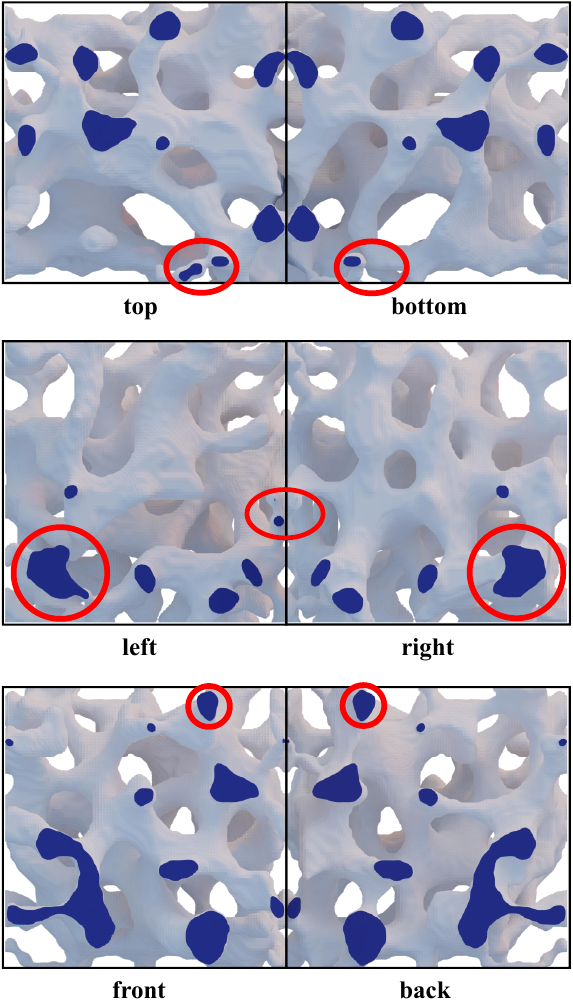}}
            \subfloat[]{
            \includegraphics[width=0.35\textwidth]{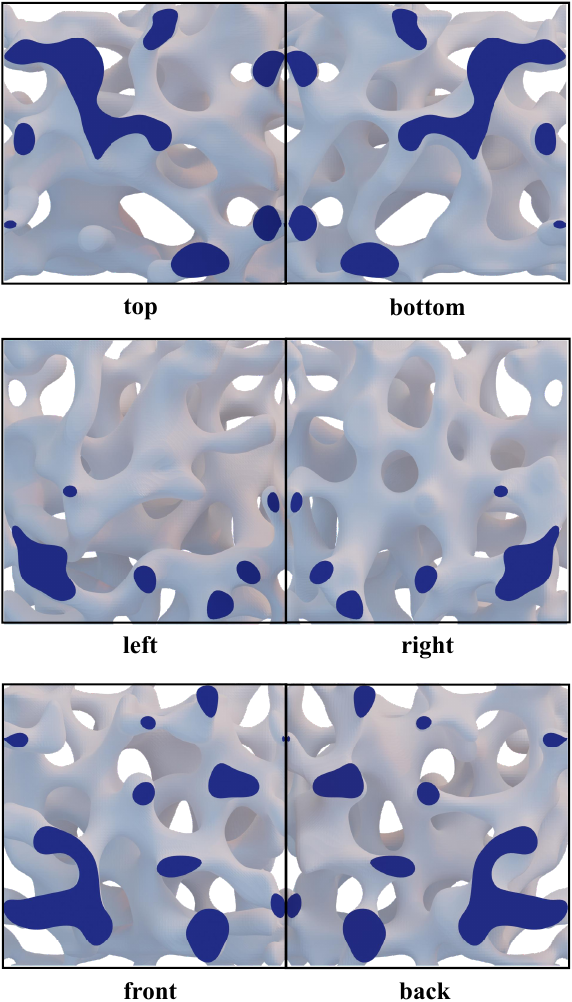}}
        \caption{Comparison with implicit unit design method in~\cite{gao2022connectivity}. The small connected components of the unit in (a) are marked in red. The six lateral faces of the units (b) and (c) are marked in blue. (a) A periodic porous sample represented by voxels. (b) The porous unit is represented by a B-spline after connectivity optimization using the method in~\cite{gao2022connectivity}. (c) The porous unit is represented by a periodic B-spline after connectivity optimization using our method. (d) Views in six directions of the unit in (b). The asymmetric regions are marked by red circles. Due to its asymmetry, the unit in (b) cannot be infinitely spliced. (e) Views in six directions of the unit in (c).}
        \label{fig:comparison}
    \end{figure}

    Given a periodic porous sample represented by voxels (see Figure~\ref{fig:comparison}(a)), we utilize the method proposed in literature~\cite{gao2022connectivity} to obtain an implicit porous unit for comparison with our proposed method.
        First, the porous sample is transformed into a discrete distance field using the DTM function. 
        Then, unlike our method, the discrete distance field is directly optimized to form a single connected component. 
        Finally, a B-spline function is used to interpolate the discrete distance field and obtain the porous structure, as shown in Figure~\ref{fig:comparison}(b). 
        Figure~\ref{fig:comparison}(d) presents the six lateral faces of this porous unit. 
      
    In our method, first, the DTM function is utilized to transform the porous sample into a discrete distance field, as described in~\cite{gao2022connectivity}. 
        Subsequently, the discrete distance field is fitted using a periodic B-spline function with a symmetric degree of $1$. 
        The resulting porous unit after connectivity optimization is shown in Figure~\ref{fig:comparison}(c). 
        Figure~\ref{fig:comparison}(e) presents the six lateral faces of this porous unit. 
    
    For clarity, we arrange the top and bottom, left and right, and front and back sides of the porous units in separate rows, forming view groups. 
        If the blue part of a view group in a row is symmetric, it indicates that when the units are spliced, the corresponding two surfaces overlap completely. 
        Furthermore, the porous units can be spliced in the corresponding direction.
        Because all the view groups in Figure~\ref{fig:comparison}(e) are symmetric, the porous unit represented by the periodic B-spline can be infinitely spliced. 
        Figure~\ref{fig:comparison}(d) displays the view groups corresponding to the unit represented by a B-spline function. 
        The view groups are asymmetric, as indicated by the red circles. 
        Although the porous sample is periodic, as shown in Figure~\ref{fig:comparison}(d), the resulting porous structure is not periodic. 
        Therefore, the porous structure obtained using the method described in~\cite{gao2022connectivity} cannot be infinitely spliced. 
        
    The main difference between these two methods lies in the fact that, first, we employ the CON-LSPIA algorithm to fit the discrete distance field while ensuring symmetry. 
        Additionally, the control coefficients of symmetric B-splines are optimized instead of optimizing the discrete distance field. 
        This ensures that the periodicity remains unchanged during the connectivity optimization process. 
        Although the method described in~\cite{gao2022connectivity} can represent complex porous structures, the size of the porous structure is fixed and depends on the size of the porous sample. 
        Designing a large-sized porous structure requires a porous sample with a large size and more computation time. 
        Additionally, the large size of porous structures increases the computational burden of simulation analysis. 
        In contrast, our method enables the representation of complex structures through unit splicing and facilitates fast simulation analysis using the homogenization methods described in Section~\ref{subsec: topology optimization}.

\subsection{Comparison of symmetric and non-symmetric periodic B-Splines}
As discussed in Section~\ref{Analysis of symmetric degree}, there is little difference in the fitting precision between symmetric periodic B-splines and non-symmetric periodic B-splines when transforming symmetric voxel samples into implicit units. 
    In this section, we further discuss the reasons for introducing symmetry.

\begin{figure}[h]
        \centering
        \includegraphics[width=0.85\textwidth]{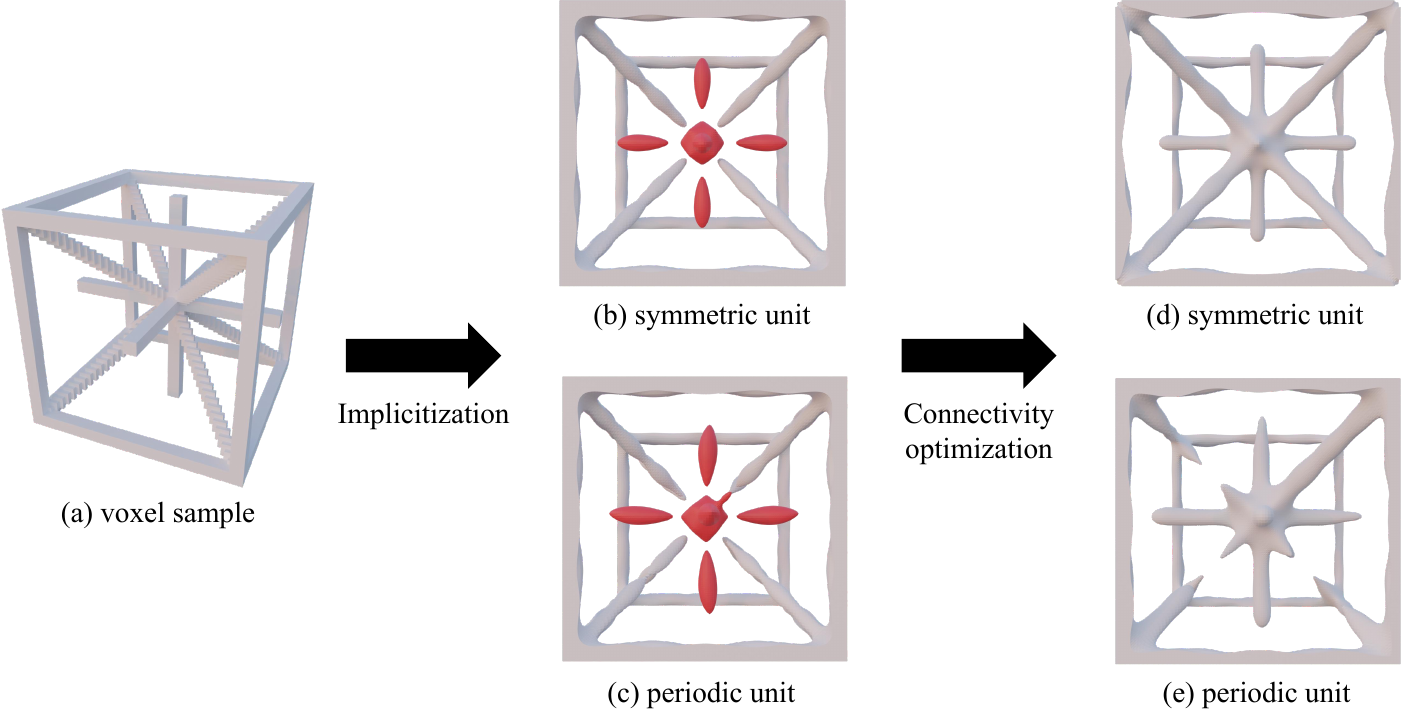}
        \caption{{Comparison of symmetric periodic B-spline and non-symmetric periodic B-spline. Isolated connected components are marked in red. (a) Artificially designed symmetric voxel sample. (b) A symmetric unit represented by a symmetric periodic B-spline. (c) A periodic unit represented by a non-symmetric periodic B-spline. (d) An optimized symmetric unit, which is connected and symmetric.  (e) An optimized periodic unit, which is connected and non-symmetric.}} 
        \label{fig:sym_vs_per}
     \end{figure}

We artificially designed a voxel sample, as shown in Figure~\ref{fig:sym_vs_per}. 
    The number of control coefficients of B-splines are chosen as $13\times 13\times 13$, and their degrees are $3\times 3\times 3$.
    The sample is then transformed into periodic B-spline functions with symmetries $r=(1,1,1)$ and $r=(6,6,6)$, corresponding to a non-symmetric and symmetric periodic function, respectively. 
    In the following discussion, we refer to them as the periodic unit and the symmetric unit for convenience. 
    The symmetric unit (Figure~\ref{fig:sym_vs_per}(b)) and the periodic unit (Figure~\ref{fig:sym_vs_per}(c)) with the same density as the voxel sample are visualized. 
    Due to the symmetry of the sample, there is no significant difference between the periodic unit and the symmetric unit. 
    To eliminate the isolated connected components (red structures in Figure~\ref{fig:sym_vs_per}(b) and (c)) introduced by implicitization, connectivity optimization described in Section~\ref{subsec: Optimization of topological features} is utilized to optimize these two units. 
    The resulting unit corresponding to symmetric periodic B-spline remains symmetric, while the resulting unit corresponding to non-symmetric periodic B-splines remains periodic. 
    Although the non-symmetric periodic B-splines represent the symmetric structure well during the fitting process, this symmetry cannot be maintained in the following connectivity optimization. 

Considering that many artificially designed units, such as the Body-Centered Cubic (BCC) and the Octet Honeycomb unit~\cite{chen2020porous}, exhibit symmetry, we need to introduce symmetric periodic B-splines to fit porous samples with symmetry. 
    This will ensure their symmetry during subsequent optimization processes.

\subsection{{Generation of 3D beam}}
\label{subsec:Topology optimization of porous model}
    
\begin{figure} [htbp]
        \centering
        \subfloat[]{
            \includegraphics[width=0.3\textwidth]{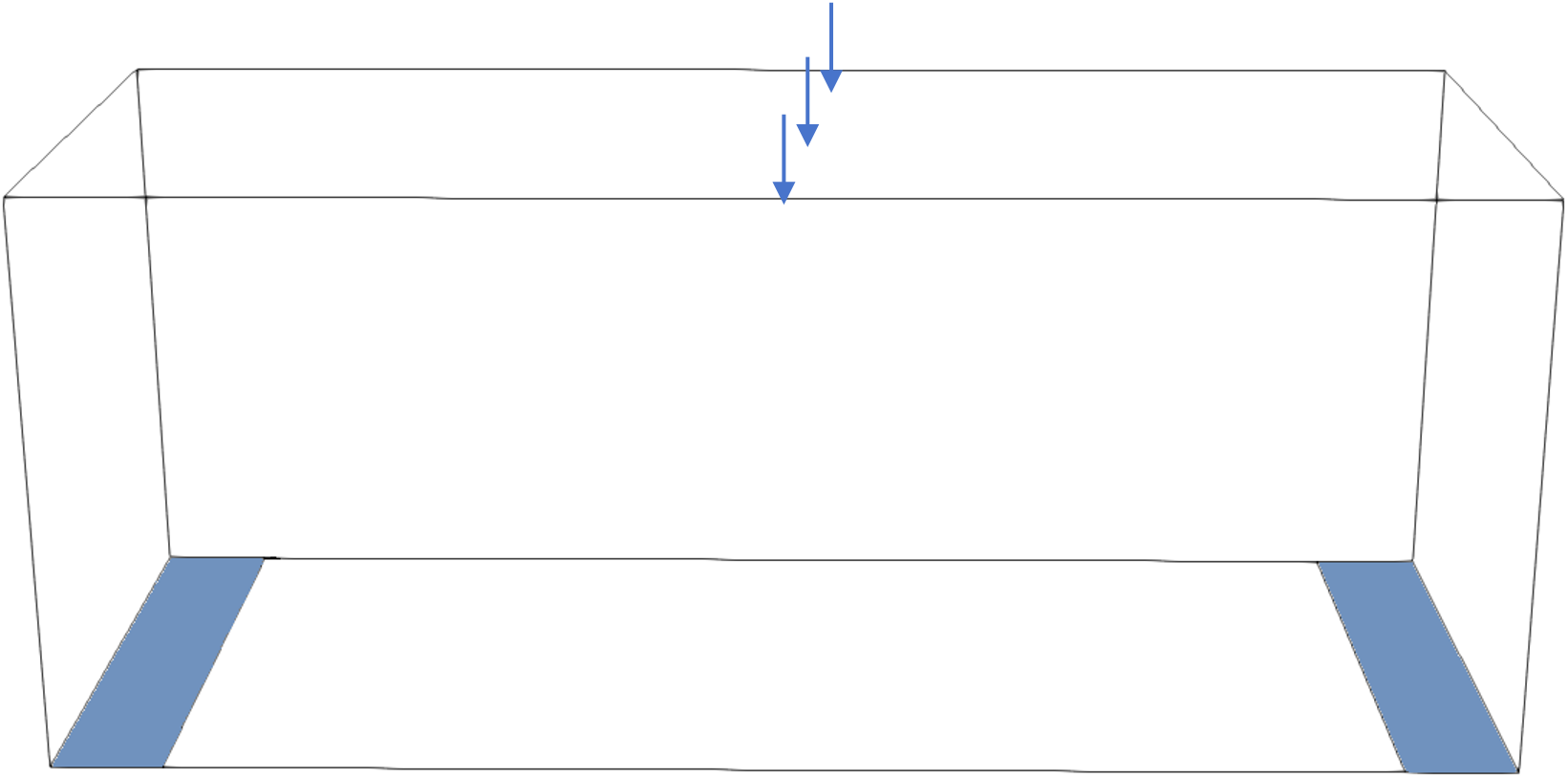}} 
        \subfloat[]{
            \includegraphics[width=0.34\textwidth]{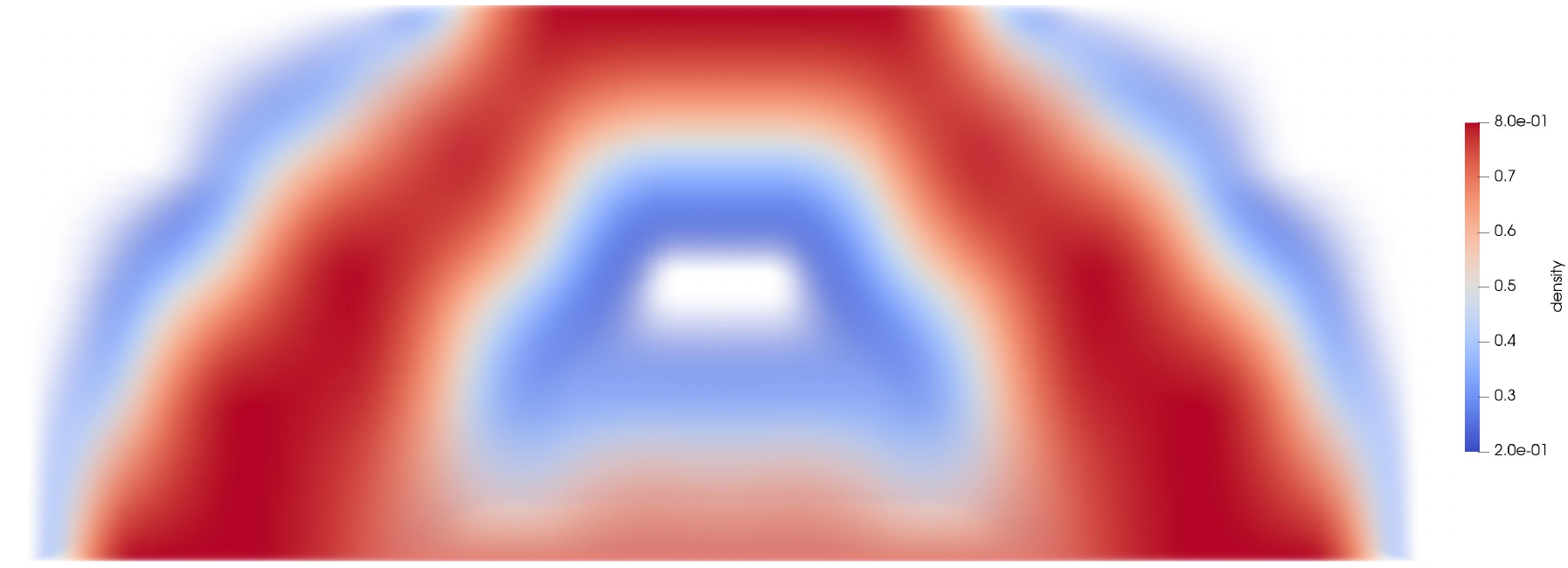}}  
        \subfloat[]{
            \includegraphics[width=0.34\textwidth]{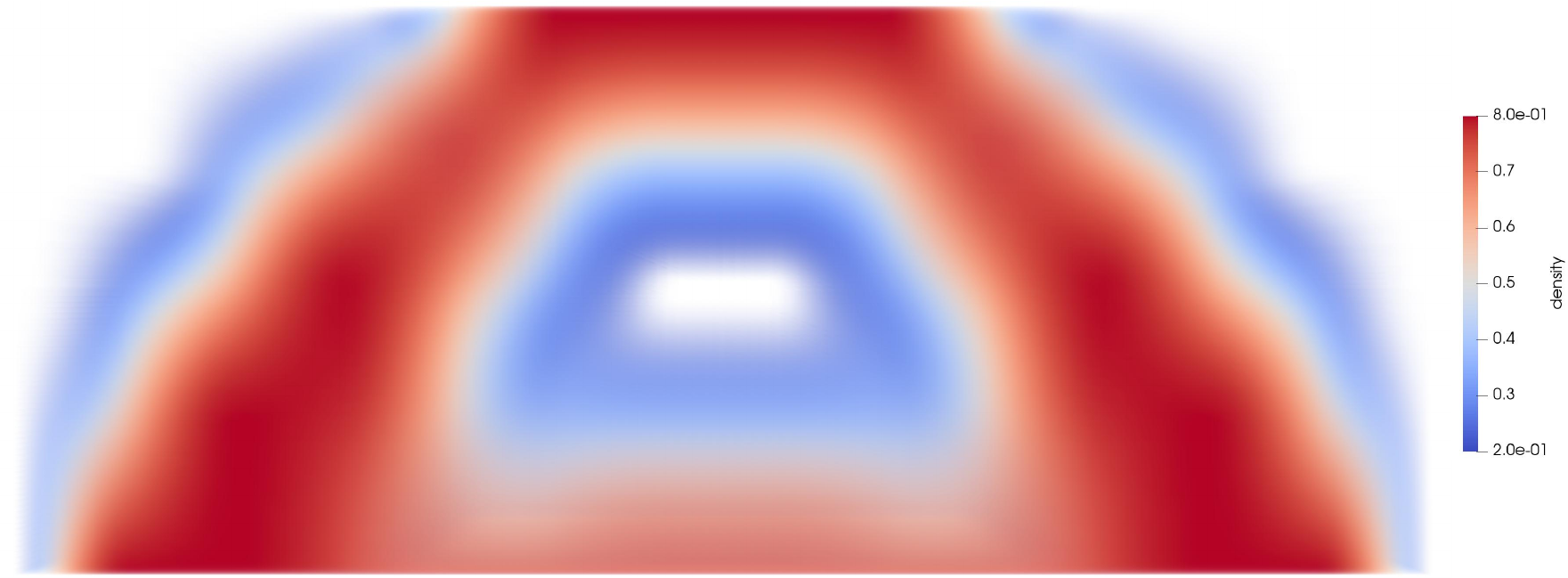}}
        \caption{Topology optimization of three-point bending test. (a) Design domain and boundary conditions. The bottom sides are fixed, and the force is uniformly applied at the center of the beam top. (b) Optimized density field of the porous unit in Table~\ref{tab:porous model}(a). (c) Optimized density field of the porous unit in Table~\ref{tab:porous model}(c).}
        \label{fig:topology optimization mechanical test} 
    \end{figure}

In this subsection, we explore the use of implicit porous units to improve model stiffness. 
        We conduct three-point bending tests to assess the effectiveness of the proposed method. 
        Given a design domain as illustrated in Figure~\ref{fig:topology optimization mechanical test}(a), the beam model is optimized using symmetric porous units from Table~\ref{tab:porous model}(a) and Table~\ref{tab:porous model}(c). 
        The unit in Table~\ref{tab:porous model}(a), designed from natural samples, and the unit in Table~\ref{tab:porous model}(c), designed artificially, are both optimized to form a connected component. 
        Topology optimization is employed to minimize compliance, with the volume constraint set to 0.4.  
        Before optimization, porous units are uniformly distributed throughout the model, as shown in the "Porous model" column of Table~\ref{tab:porous model}(b) and Table~\ref{tab:porous model}(d). 
        The optimized density fields of these two units are depicted in Figure~\ref{fig:topology optimization mechanical test}(a) and Figure~\ref{fig:topology optimization mechanical test}(c), respectively. 
        Finally, the threshold distribution field of the porous unit is obtained to align with the density field, and the optimized model is depicted in the "Porous model" column of Table~\ref{tab:porous model}(a) and Table~\ref{tab:porous model}(c), respectively. 
    
\begin{figure} [htbp]
        \centering
        \subfloat[]{
            \includegraphics[width=0.35\textwidth]{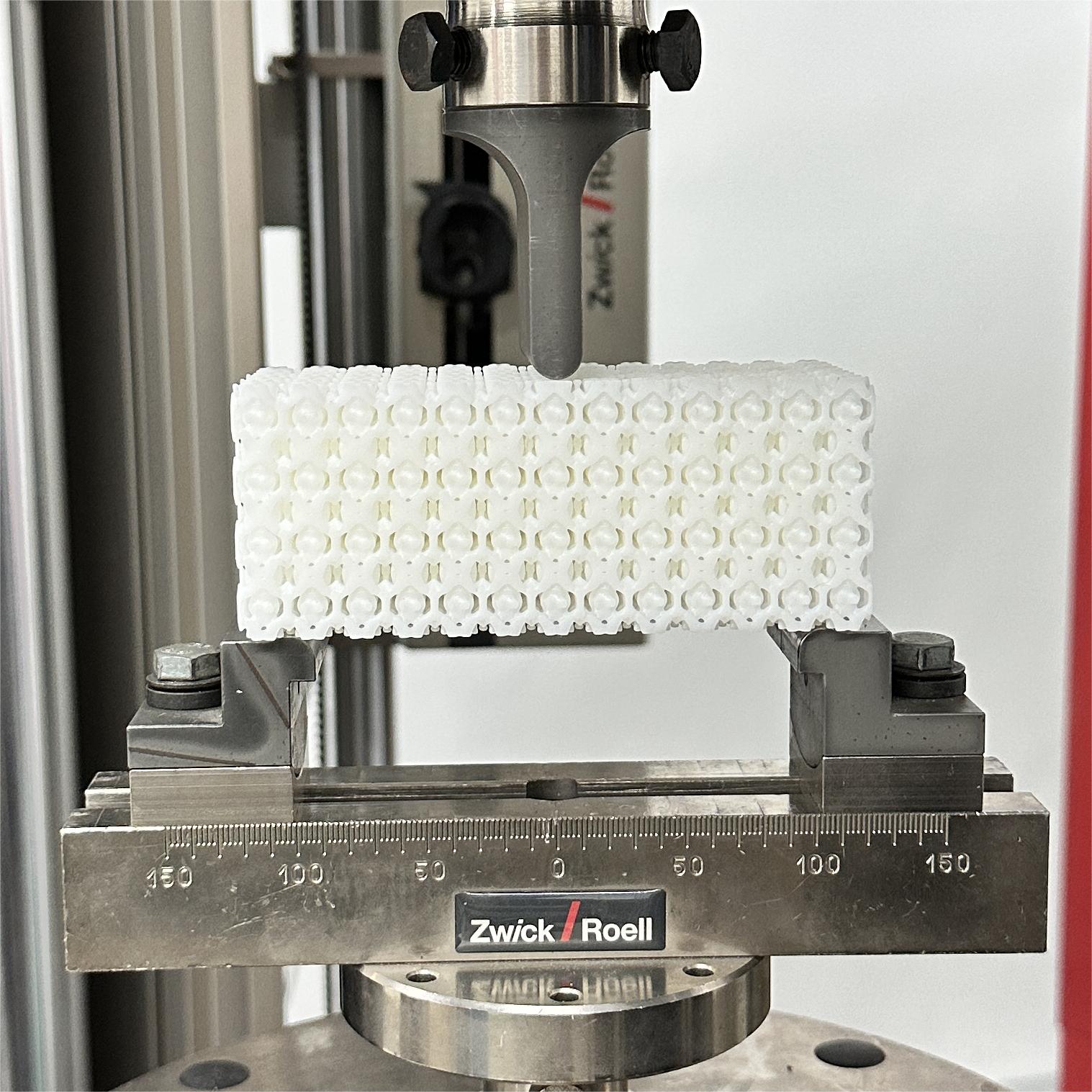}} 
        \subfloat[]{
            \includegraphics[width=0.5\textwidth]{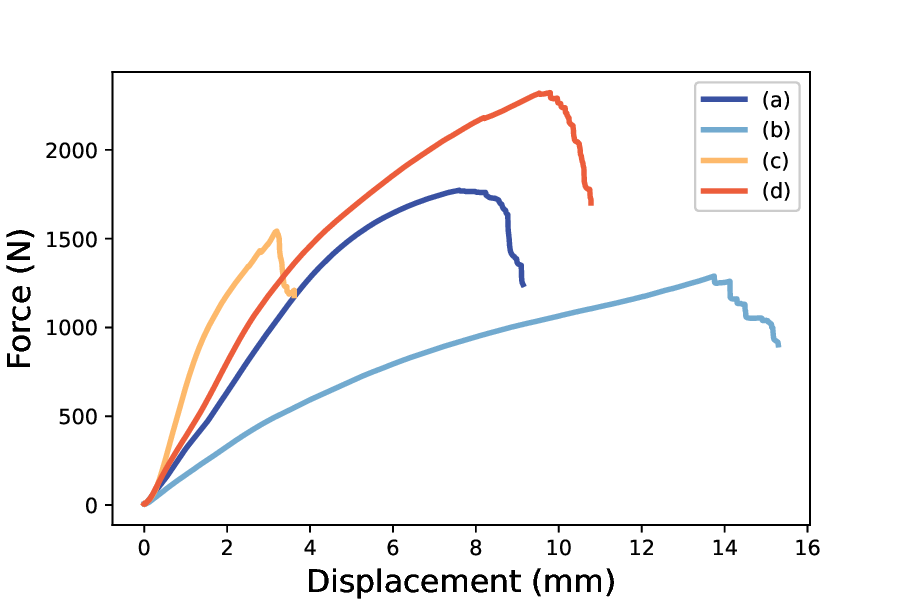}}  
        \caption{The experimental equipment and experimental results. (a) The equipment used for testing three-point bending. (b) Corresponding displacement-force curves for the four porous models are shown in Table~\ref{tab:porous model}.}
        \label{fig:mechanical test} 
\end{figure}

    To further demonstrate the effectiveness of the topology optimization results, we manufactured porous models using additive manufacturing and performed a mechanical test. 
        We utilized additive manufacturing to produce optimized and uniform porous models with the size of $120\times 48\times 48 $ $mm^3$. 
        The corresponding uniform porous model and optimized porous model for these two units are shown in the "Manufactured porous model" column of Table~\ref{tab:porous model}. 
        The three-point bending test was performed using the equipment shown in Figure~\ref{fig:mechanical test}(a). 
        Curves (a) and (b) in Figure~\ref{fig:mechanical test}(b) correspond to the porous models of Table~\ref{tab:porous model}(a) and Table~\ref{tab:porous model}(b), respectively. Curves (c) and (d) in Figure~\ref{fig:mechanical test}(b) correspond to the porous models of Table~\ref{tab:porous model}(c) and Table~\ref{tab:porous model}(d), respectively.
        It can be observed that the compliance of the porous model decreases after topology optimization (lower displacement under the same external force).
        This clearly demonstrates the effectiveness of the optimization method proposed in this study. 
        Furthermore, both the porous samples and artificially designed porous units can be applied to enhance the model stiffness. 
        Therefore, the new representation of the porous unit shows great potential for application. 
    
    \begin{table}[h]
        \centering
        \caption{Uniform and optimized porous models with the volume ratio of 0.4 and their manufactured models.}
        \label{tab:porous model}
        \begin{tabular}{ | c | c | c | c | }
        \hline
            & Implicit unit & Porous model & Manufactured porous model\\ \hline
        (a) & 
        \begin{minipage}[b]{0.15\linewidth}
                  \centering
                  \raisebox{-.5\height}{\includegraphics[width=\linewidth]{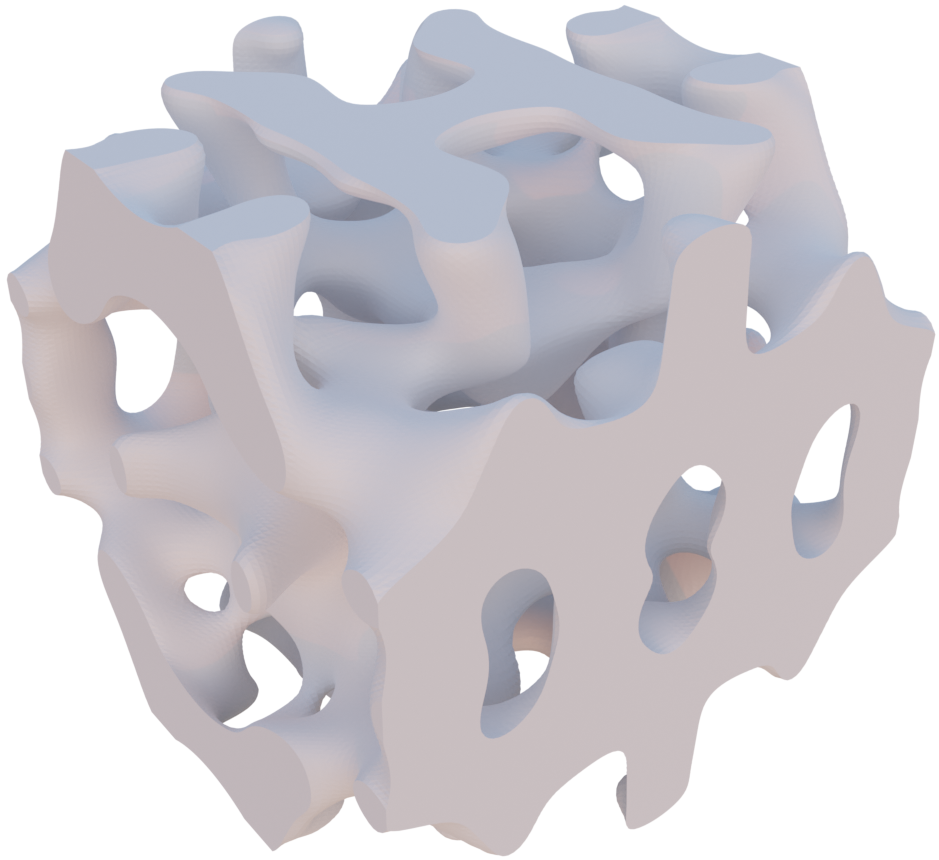}}
              \end{minipage}
        &
        \begin{minipage}[b]{0.37\linewidth}
                  \centering
                  \raisebox{-.5\height}{\includegraphics[width=\linewidth]{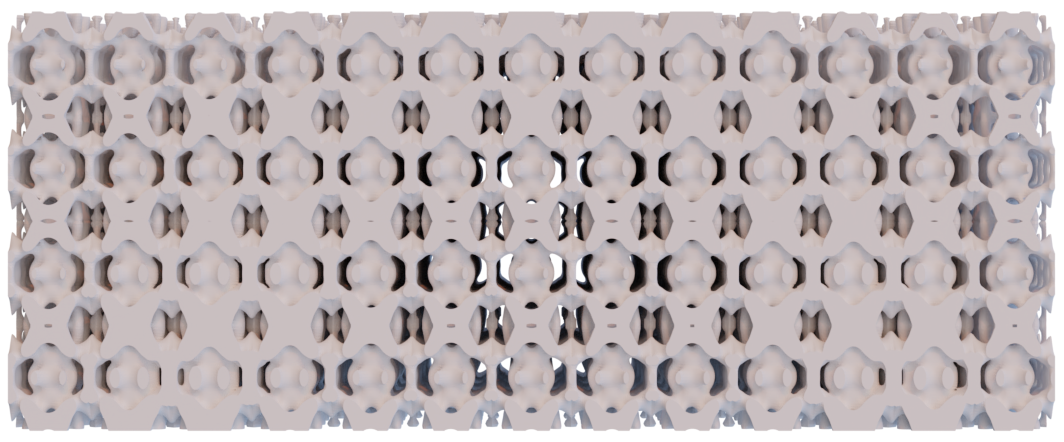}}
              \end{minipage}
              & \begin{minipage}[b]{0.37\linewidth}
                \centering
                \raisebox{-.5\height}{\includegraphics[width=\linewidth]{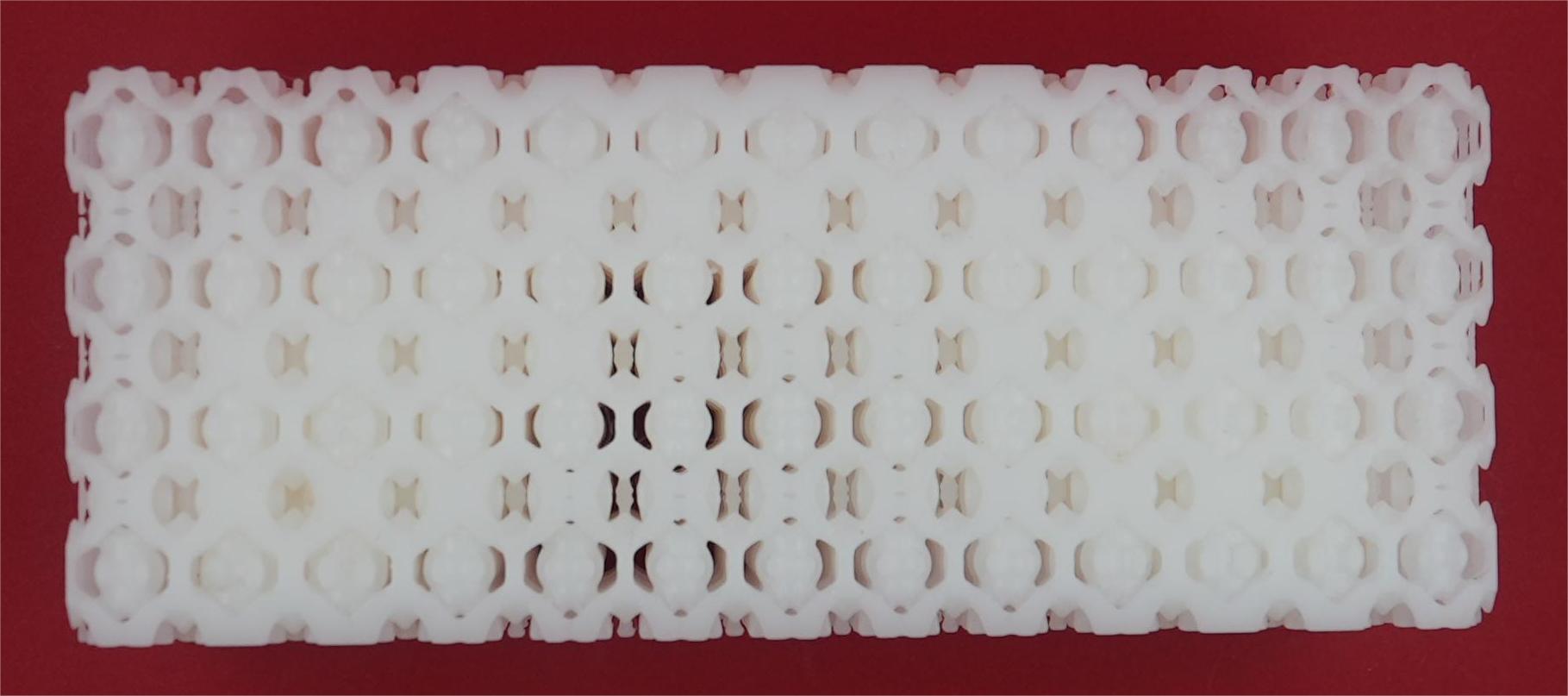}}
            \end{minipage}
              \\ \hline
        (b) &
        \begin{minipage}[b]{0.15\linewidth}
                  \centering
                  \raisebox{-.5\height}{\includegraphics[width=\linewidth]{10_opt_0.3.png}}
              \end{minipage}
        &
        \begin{minipage}[b]{0.37\linewidth}
                  \centering
                  \raisebox{-.5\height}{\includegraphics[width=\linewidth]{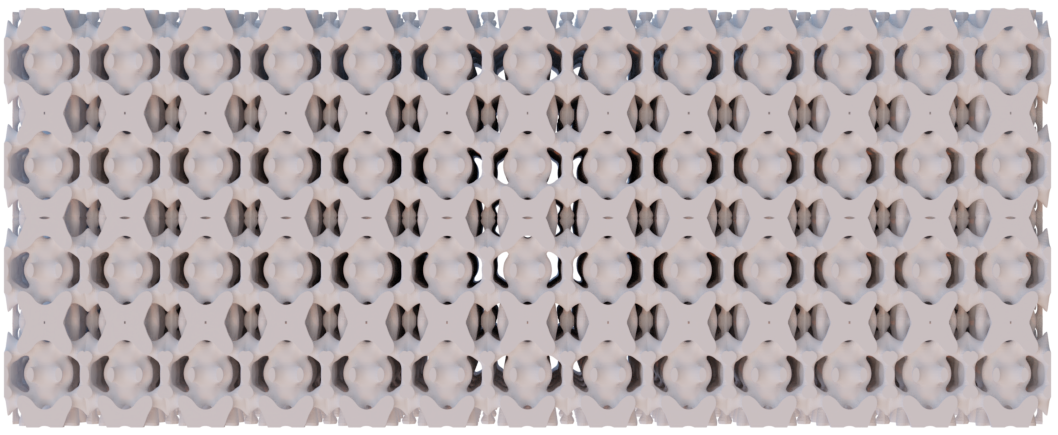}}
              \end{minipage}
              & \begin{minipage}[b]{0.37\linewidth}
                \centering
                \raisebox{-.5\height}{\includegraphics[width=\linewidth]{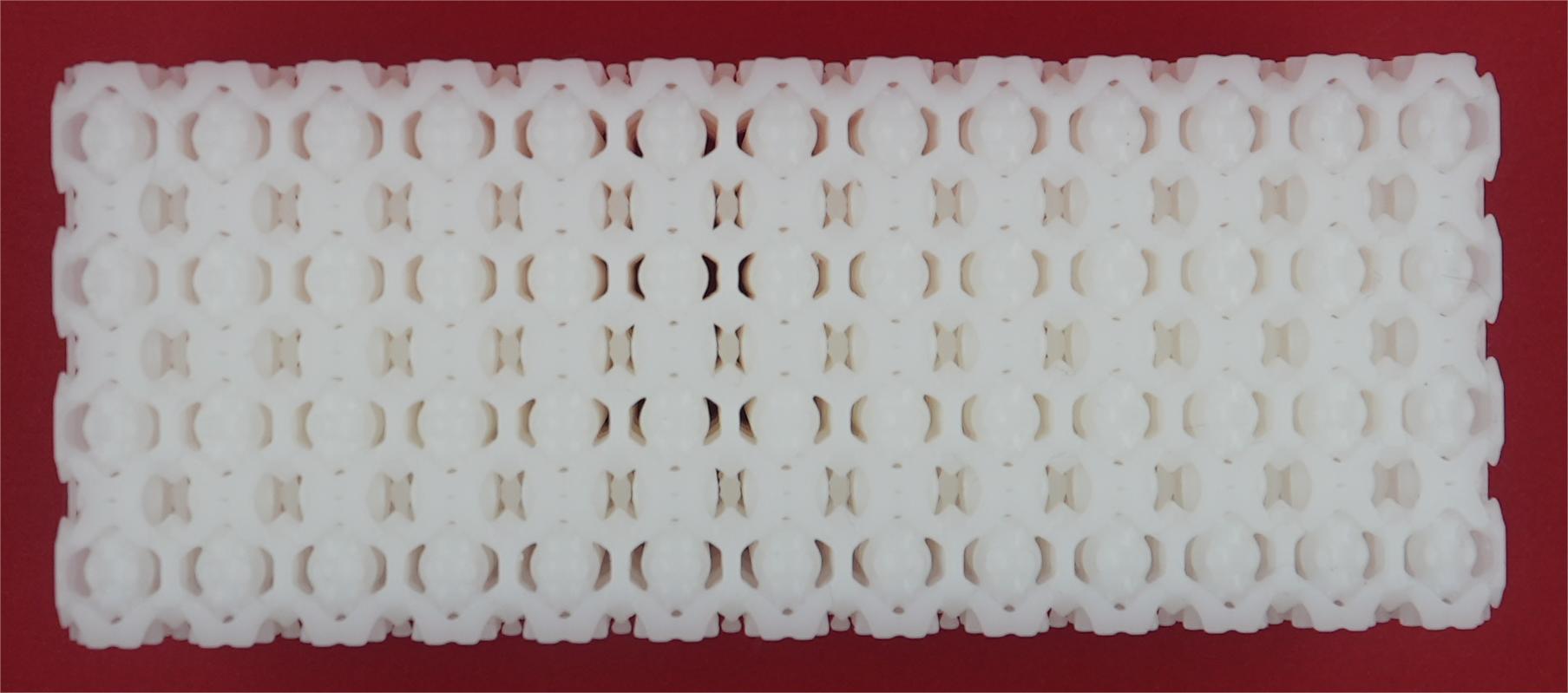}}
            \end{minipage}
              \\ \hline
        (c) &
        \begin{minipage}[b]{0.15\linewidth}
                  \centering
                  \raisebox{-.5\height}{\includegraphics[width=\linewidth]{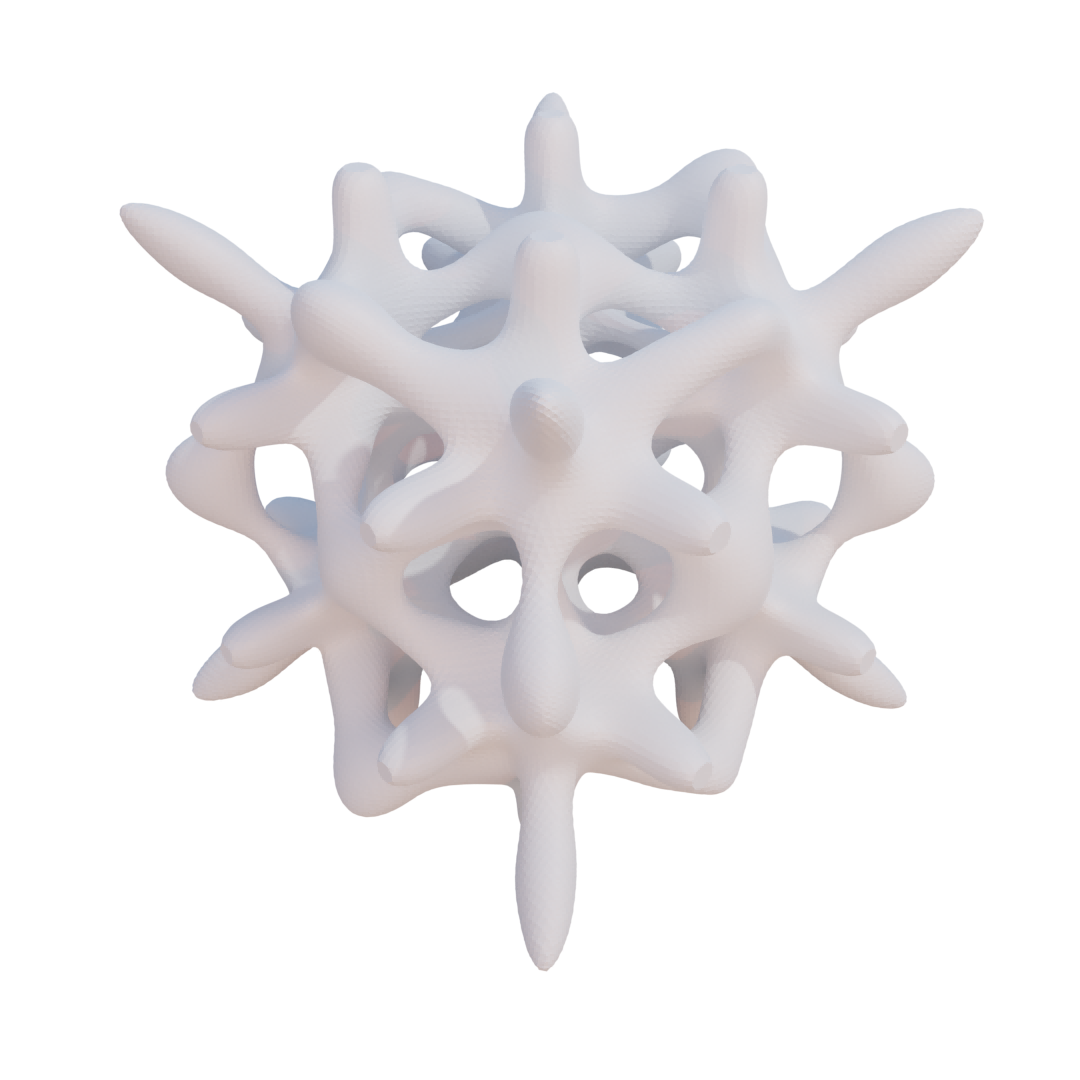}}
              \end{minipage}
        &
        \begin{minipage}[b]{0.37\linewidth}
                  \centering
                  \raisebox{-.5\height}{\includegraphics[width=\linewidth]{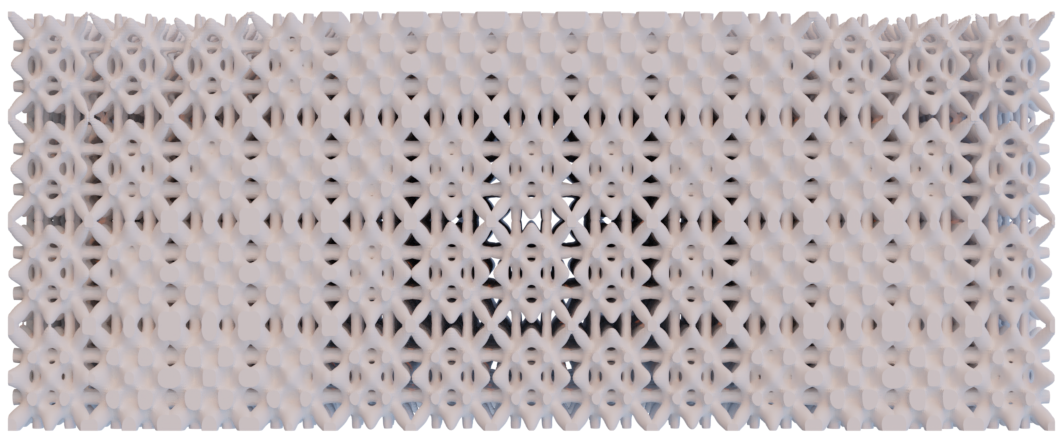}}
              \end{minipage}
              & \begin{minipage}[b]{0.37\linewidth}
                \centering
                \raisebox{-.5\height}{\includegraphics[width=\linewidth]{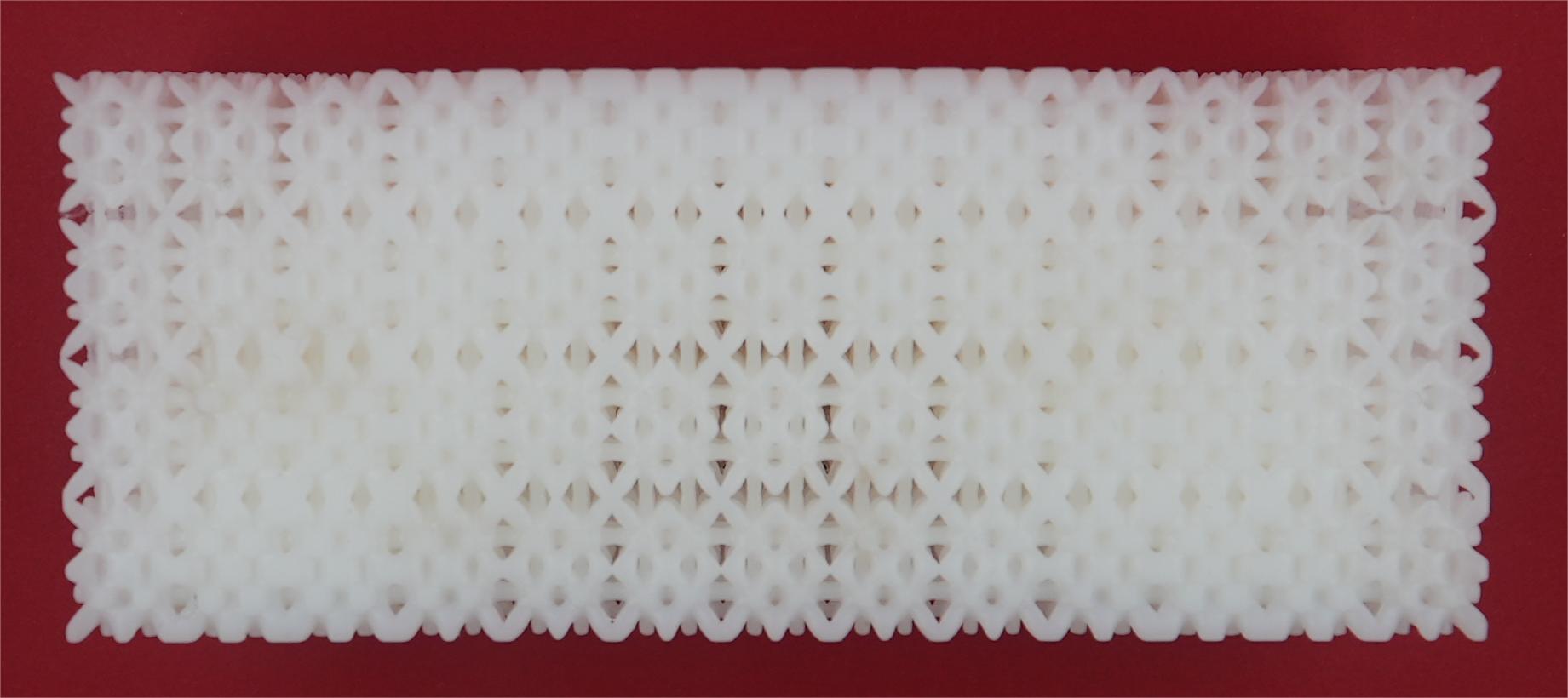}}
            \end{minipage}
              \\ \hline
        (d) &
        \begin{minipage}[b]{0.15\linewidth}
                  \centering
                  \raisebox{-.5\height}{\includegraphics[width=\linewidth]{artificial_after_opt.png}}
              \end{minipage}
        &
        \begin{minipage}[b]{0.37\linewidth}
                  \centering
                  \raisebox{-.5\height}{\includegraphics[width=\linewidth]{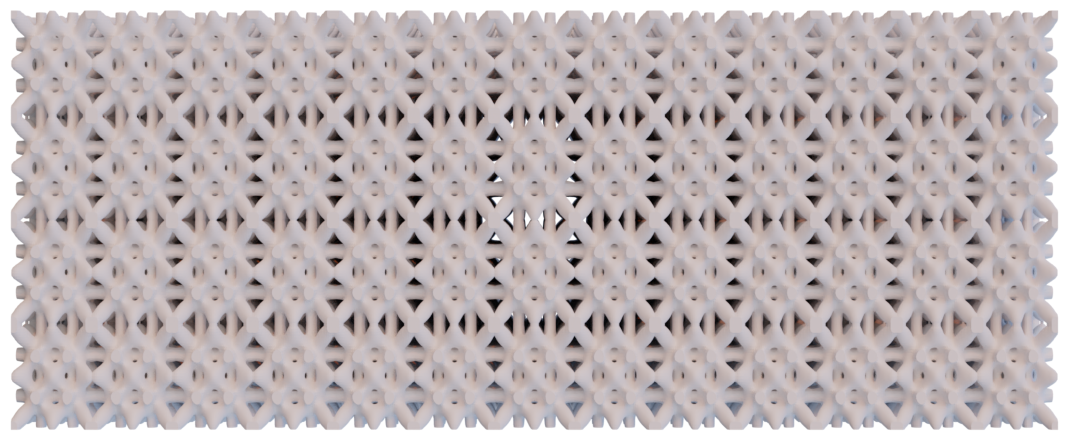}}
              \end{minipage}
              & \begin{minipage}[b]{0.37\linewidth}
                \centering
                \raisebox{-.5\height}{\includegraphics[width=\linewidth]{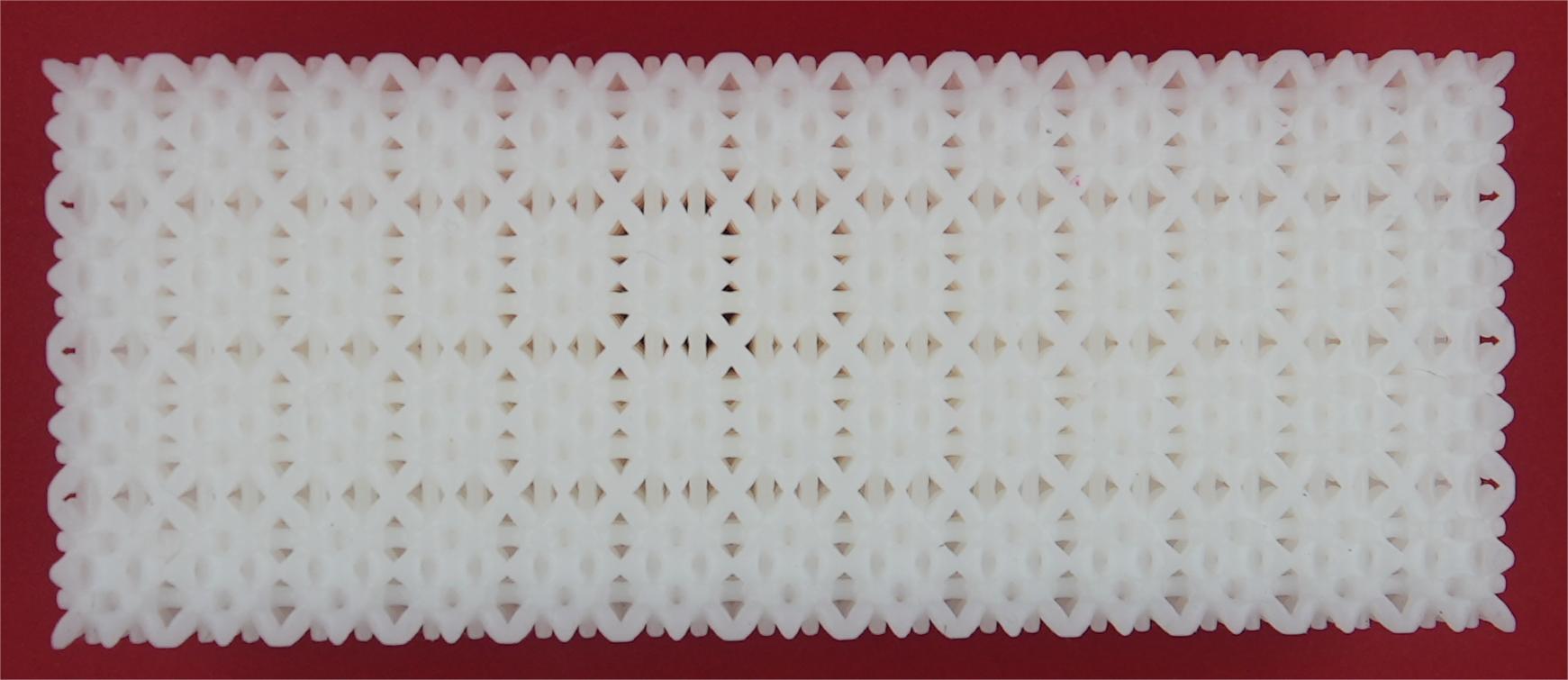}}
            \end{minipage}
              \\ \hline
            \end{tabular}
            \vspace{5pt}

    \end{table}   
    
\subsection{Generation of general model}
\begin{figure}[h]
      \centering
      \includegraphics[width=0.95\textwidth]{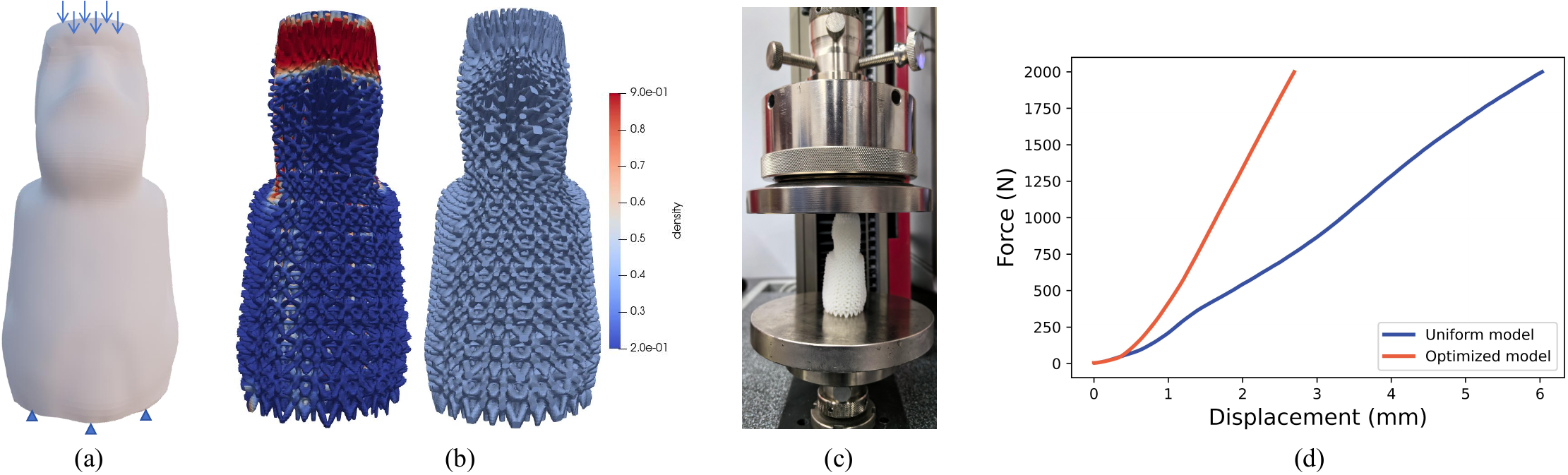}
      \caption{{Compression experiment. Porous models with a volume fraction of 0.4. (a) Design domain and boundary conditions. The bottom of the model is fixed, and the top is subjected to a uniformly force. (b) Porous models. The color corresponds to the relative density of the porous model. The model on the left is obtained after topology optimization. The model on the right is a uniform model. (c) The experimential equipment. (d) The displacement-force curves.  }} 
      \label{fig:compression model} 
   \end{figure}

We conduct a compression test based on the model shown in Figure~\ref{fig:compression model}(a) as the design domain. 
   The bottom of the model is fixed, and a uniform force is applied from the top. 
   The uniform model before topology optimization and the model obtained after topology optimization are shown in Figure~\ref{fig:compression model}(b). 
   The color of the model reflects its relative density. 
   The topology optimization method changes the density distribution of the model to improve its stiffness. 
   Subsequently, the porous model is manufactured using additive manufacturing and subjected to compression testing (see Figure~\ref{fig:compression model}(c)). 
   As illustrated in Figure~\ref{fig:compression model}(d), the displacement of the optimized model is smaller than the uniform one under the same external force. 
   Therefore, the optimized model exhibits smaller compliance, that is, higher stiffness.

\subsection{Application}
One potential application of this study is in bone tissue engineering, which is typical biological applications of porous structures~\cite{feng2022triply}. 
   Porous scaffolds are used as substitutes for defective bone tissue to provide necessary mechanical support. 
   Additionally, porous scaffolds have good biocompatibility due to their high porosity, which aids in cell proliferation~\cite{gao2021porous}.

\begin{figure}[h]
      \centering
      \includegraphics[width=0.95\textwidth]{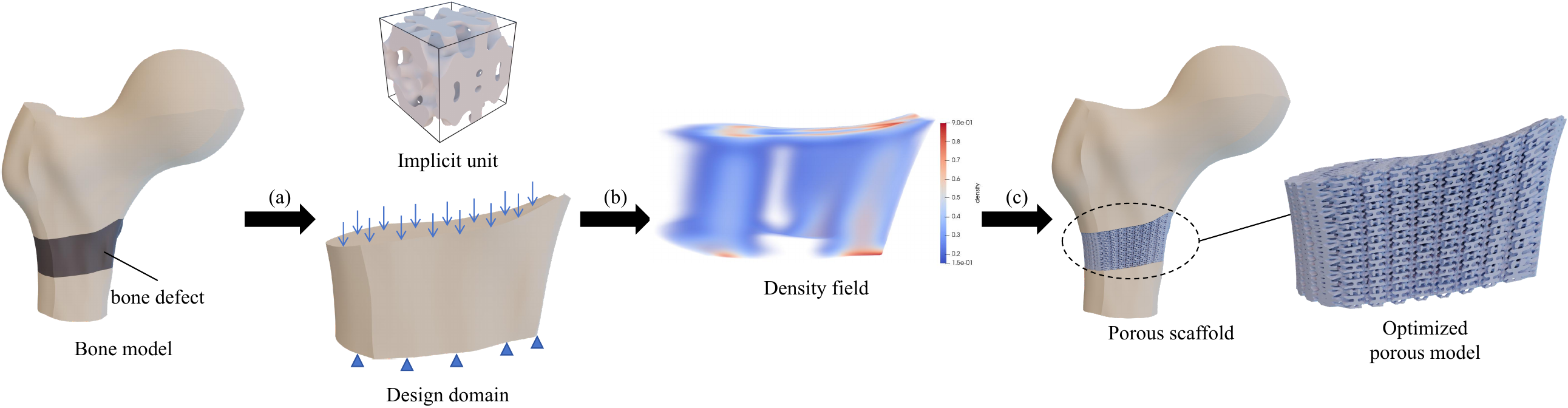}
      \caption{{Generation of bone scaffold. (a) Given a bone defect region, the design domain and basic implicit unit are obtained based on the requirement of applications. 
      (b) The optimal density field distribution is obtained using topology optimization method. 
      (c) A porous model conforming to the density field is generated. It is then combined with the bone model to form a bone scaffold.} } 
      \label{fig:bone_scaffold} 
   \end{figure}

As shown in Figure~\ref{fig:bone_scaffold}, given a local bone model and a defect region, the proposed method can customize bone scaffolds for a patient. 
   By specifying the basic implicit unit and the force and displacement conditions of the porous model, the topology optimization method introduced in Section~\ref{subsec: topology optimization} can generate a density distribution that maximizes stiffness to ensure the mechanical property of the scaffold. During the optimization process, we set the volume constraint of the model to 0.25. 
   Considering the loading pattern of the bone, we assume that the model is subjected to a uniformly distributed downward force on the upper surface, while the lower surface is fixed. 
   After the topology optimization, the relative compliance of the model (the compliance of the porous model divided by the compliance of the solid model) decreases from 11.77 to 9.59. 
   This indicates an increase in the stiffness of the model.
   Subsequently, the porous model can be fabricated using additive manufacturing and assembled into the original bone model to obtain the bone scaffold.

It should be noted that this study proposes a method to convert natural voxel samples into implicit units.  
   Due to the limitations of CT scanning, it is difficult to completely capture and accurately manufacture the microstructure of defective bones. 
   Instead, CT technology can be used to scan the local microstructure samples of the patient's bone. 
   Subsequently, these local samples can be transformed into an implicit periodic representation using the method proposed in this study. 
   The experiment in Section~\ref{Analysis of symmetric degree} has already shown that the introduction of periodicity does not significantly affect the structure of the samples. 
   Moreover, introducing periodicity allows these small-scale samples to be infinitely spliced to any size to form a bone scaffold. 
   Thus, the designed bone scaffold will have a similar microstructure configuration to that of the patient, aiming to achieve the desired mechanical properties and biocompatibility.

\subsection{Discussion}
We use the periodic B-spline function as a new representation for implicit porous units, allowing them to be infinitely spliced throughout the entire space. 
    At the same time, our method can transform voxel porous samples into implicit porous units. 
    Since voxel porous samples are more user-friendly for interactive design and acquisition compared to implicit porous units, users can more easily design implicit porous units. 
    The experiments in this section effectively demonstrate that choosing periodic B-splines with a symmetry degree of $r>1$ for fitting can effectively obtain periodic implicit porous units for arbitrary voxel porous samples. 
    Furthermore, the symmetry degree $r$ of the implicit porous units remains unchanged during connectivity optimization. 
    Moreover, we apply the implicit porous units under new representation to the problem of minimizing the compliance of porous models to showcase their potential applications.

This method still has some limitations.
On one hand, for voxel porous samples of sizes, $100$ and $1000$, different numbers of control coefficients need to be used for fitting in order to achieve the desired fitting precision. 
    Therefore, for voxel porous datasets of different sizes, an algorithm is needed to automatically select appropriate numbers of control coefficients for fitting. 
    On the other hand, the choice of symmetry degree $r$ significantly affects the fitting precision. 
    For a non-periodic voxel porous sample, setting $r=(1,1,1)$ can obtain a periodic implicit porous unit without significantly changing its morphology. 
    For a symmetric voxel porous sample, fitting with periodic B-splines of $r=(1,1,1)$ and $r=(\lfloor {\frac{n_u}{2}} \rfloor,\lfloor {\frac{n_v}{2}} \rfloor,\lfloor {\frac{n_w}{2}} \rfloor)$ can both achieve high fitting precision. 
    However, the periodic B-spline corresponding to $r=(1,1,1)$ cannot maintain the inherent symmetry of the symmetric sample during subsequent connectivity optimization, like the $r=(\lfloor {\frac{n_u}{2}} \rfloor,\lfloor {\frac{n_v}{2}} \rfloor,\lfloor {\frac{n_w}{2}} \rfloor)$ situation. 
    Therefore, future work needs to propose methods to analyze the symmetry of input voxel samples to select the optimal symmetry degree $r$.

    \section{Conclusion}
    \label{sec: Conclusion}
    In this study, we propose a design method for implicit porous units using periodic B-spline functions as a new representation. 
        The implicit unit is obtained from a voxel porous sample. 
        Firstly, the discrete distance field of the porous sample is calculated using the DTM field to minimize the influence of noise. 
        Next, we use the proposed iterative fitting method (CON-LSPIA) to obtain the periodic B-spline that represents a porous unit. 
        Subsequently, the connectivity of the porous unit can be optimized by adjusting the control coefficients. 
        Afterward, the porous unit can be applied to topology optimization to enhance the stiffness of the porous model. 
        The mechanical test suggests that the designed implicit units can enhance the stiffness of the porous model. 
    
    CT technology provides a large number of 3D binary images of porous structures. 
        The method proposed in this study can convert these images into a database of implicit porous units. 
        Retrieving porous units from the database that meet the expected physical performance poses an interesting research challenge. 
        Moreover, multi-unit topology optimization is gaining significant attention in recent research. 
        The purpose of multi-unit topology optimization is to select different porous units from a database to form heterogeneous porous structures, aiming to improve model performance. 
        Employing the porous unit database for multi-unit topology optimization is a future work for us.

\section*{Acknowledgments}
\noindent
This work is supported by National Natural Science Foundation of China under Grant nos. 62272406 and 61932018.

\section*{Declaration of competing interest}
The authors declare that they have no known competing financial interests or personal relationships that could have appeared to influence the work reported in this paper.

\section*{Declaration of Generative AI and AI-assisted technologies in the writing process}
During the preparation of this work the author(s) used ChatGPT in order to improve language and readability. After using this tool/service, the author(s) reviewed and edited the content as needed and take(s) full responsibility for the content of the publication.

\bibliographystyle{elsarticle-num} 
\bibliography{mybibfile}

\end{document}